%% file: ms.tex
\documentclass[journal]{IEEEtran}

\pdfoutput=1 
\usepackage{pgfplots}
\usepackage{tikz}
\usepackage{circuitikz}
\usepackage{graphicx}
\usepackage{amsmath} 
\usepackage{mathtools} 
\usepackage{miller} 

\usepackage[caption=false]{subfig}	
\pgfplotsset{compat=1.16,tick label style={font=\small}}

\begin{document}

\title{Wave Propagation in Electric Periodic Structure in Space with Modulation in Time (2D+1). I. Theory}%

\author{José~de~Jesús~Salazar-Arrieta,
		Peter~Halevi%
\thanks{This work was supported by the CONACyT under Grant A1-S-45628 \textit{(Corresponding author: José~de~Jesús~Salazar-Arrieta)}}%
\thanks{José~de~Jesús~Salazar-Arrieta was with the Department of Electronics, Instituto Nacional de Astrofísica, Óptica y Electrónica, Tonantzintla, 72000 Puebla, Mexico, e-mail: (jose.arrieta@inaoep.mx).}%
\thanks{Peter~Halevi is with the Department of Electronics, Instituto Nacional de Astrofísica, Óptica y Electrónica, Tonantzintla, 72000 Puebla, Mexico, e-mail: (halevi@inaoep.mx).}}

\IEEEaftertitletext{\vspace{-1\baselineskip}}
\maketitle

\begin{abstract}
We studied electromagnetic wave propagation in a system that is periodic in both space and time, namely a discrete 2D transmission line (TL) with capacitors modulated in tandem externally. Kirchhoff's laws lead to an eigenvalue equation whose solutions yield a band structure (BS) for the circular frequency $\omega$ as function of the phase advances $k_{x}a$ and $k_{y}a$ in the plane of the TL. The surfaces $\omega(k_{x}a, k_{y}a)$ display exotic behavior like forbidden $\omega$ bands, forbidden $k$ bands, both, or neither. Certain critical combinations of the modulation strength $m_{c}$ and the modulation frequency $\Omega$ mark transitions from $\omega$ stop bands to forbidden $k$ bands, corresponding to phase transitions from no propagation to propagation of waves. Such behavior is found invariably at the high symmetry $\mathbf{X}$ and $\mathbf{M}$ points of the spatial Brillouin zone (BZ) and at the boundary $\omega=(1/2)\Omega$ of the temporal BZ. At such boundaries the $\omega(k_{x}a, k_{y}a)$ surfaces in neighboring BZs assume conical forms that just touch, resembling a South American toy ``diábolo''; the point of contact is thus called a ``diabolic point''. Our investigation reveals interesting interplay between geometry, critical points, and phase transitions.
\end{abstract}

\input{SecIntroduction} 
\input{SecEigenValueProblem}
	\input{SubSecCaseStatic}
	\input{SubSecCaseEmptyRed}
\input{SecBandsStructure}
	\input{SubSecDir10_11}
	\input{SubSecArbitraryDir}
\input{SecDiabolicPoints}
\input{SecConclusion}

\ifCLASSOPTIONcaptionsoff
 \newpage
\fi
\IEEEtriggeratref{20}
\bibliographystyle{IEEEtran}
\bibliography{BibReference}

\end{document}

%% file: SecIntroduction.tex
\section{Introduction.}

\begin{figure*}[!t] 
\captionsetup[subfigure]{labelformat=empty}
\centering
	\subfloat[\label{Cto2D}]{\input{SubFigura1a}}
	\qquad
	\subfloat[\label{Celda2D}]{\input{SubFigura1b}}
\caption{(a) 2D transmission line or temporal electric crystal in the $x-y$ plane with square lattice. (b) A unit cell with constant inductances L and periodically modulated capacitances $C(t)$. This is a discrete or granular lattice with each unit cell specified by a pair of integers $(x, y) = (p, q)$. The corresponding currents $I_{p,q}^{x,y}$ and voltages $V_{p,q}$ are indicated.}
\label{Cto_Celda_2D}
\end{figure*}
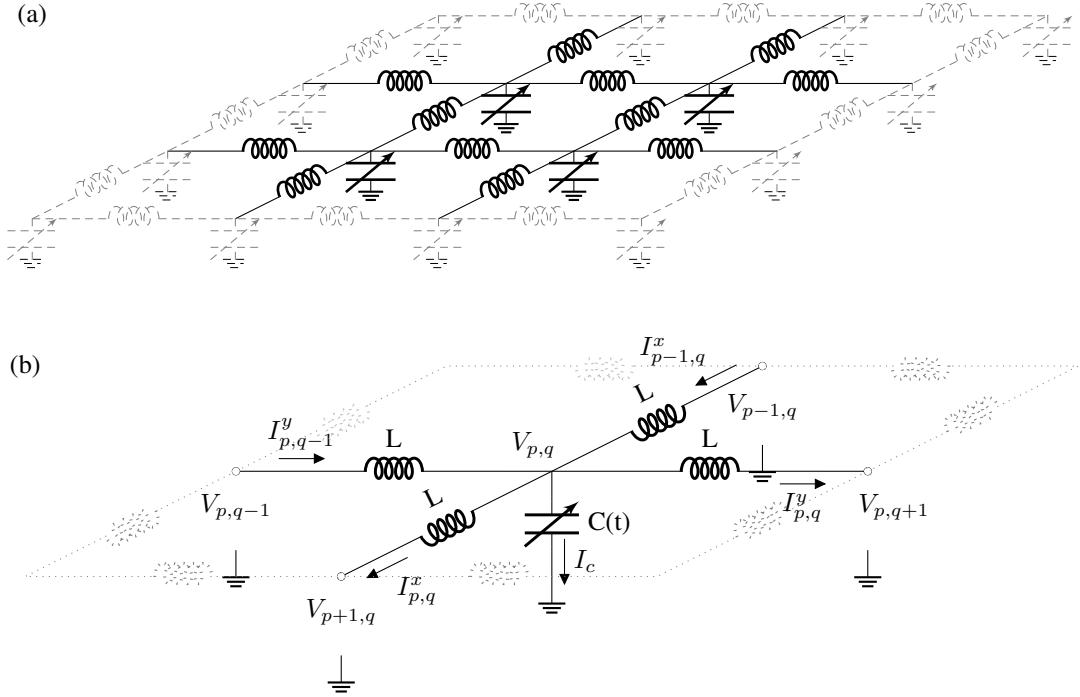

In this paper we explore electromagnetic wave (EMW) propagation in a system that is spatially periodic in a plane and additionally, also periodic in time. Concretely, our system is a two-dimensional (2D) discrete (``granular'') transmission line (TL) with capacitors (varactors) that are periodically modulated in time, Fig.\ref{Cto_Celda_2D}. Part II of this work will be dedicated to the corresponding simulations.

Wave propagation in TLs and crystals has been amply studied in one, two, and three spatial dimensions in \cite{lbrillouin}, the 3D case of crystals being basic to Solid State physics \cite{Kittel}. The band structures of these crystals depend on intrinsic properties that are difficult to alter. On the other hand, artificial structures with 1D, 2D, and 3D periodicities, known as ``photonic crystals'', were developed and are much more easily tuned \cite{joanopulos}; with appropriate design and long wavelengths in comparison to the unit cell, these can even display ``metamaterial'' behavior with backward propagating waves \cite{caloz}. Effects of the periodicity have been also researched in detail in 1D TLs in \cite{Pozar}. The band structures of all the aforementioned systems are periodic in the reciprocal lattice and they display forbidden frequency bands. In a 2D low-pass TL, Afshari et al \cite{RelDisp_Estatica} demonstrated that a high quality spatial filter, called ``electric prism'', can be designed. The same group, using nonlinear voltage dependent varactors, showed that it is possible to modify the spectral content, width, and height of a pulse \cite{WideSing}. They also performed more detailed studies of generation of harmonics with selective maximization of their amplitudes \cite{HarmonGen}, of nonlinear constructive interference of voltage waves \cite{ConsInter}, and of a high-speed quantifier \cite{Cuantizador}. Further, Afshari's group implemented an integrated circuit in CMOS technology so as to improve the amplitude and width of an incident signal \cite{cmos}. On the other hand, with merely interchanging the capacitors and inductors (namely, the dual structure of the low-pass TL) a 2D TL can exhibit ``left-handed'' behavior for long wavelengths, giving rise to negative refraction \cite{NegSuport2d}. Eleftheriades's group combined a low-pass TL and its dual, thus obtaining a composite right/left handed (CRLH) medium with planar negative refractive index \cite{NegLoaded, NegMult}. It is also possible to achieve negative refraction by means of inductive coupling of 2D TLs \cite{NegHUriel} and the same effect can as well be realized by an anisotropic arrangement of capacitors and inductors \cite{NegHiperbolic}.
 
The parameters of the TLs and crystals can also depend on time, in the case of crystals, they were called ``temporal photonic crystals'' \cite{Zurita}. Zurita and Halevi researched the photonic band structure of a uniform, modulated dielectric, and the optical response of a slab of such a medium \cite{Zurita}. They also reported parametric resonances in a time-periodic slab for critical widths of its thickness \cite{ZuritaResonancias}. These investigations were generalized to allow for modulation of the permeability, in addition to the permittivity \cite{Sabino, Sabino2018}. The hallmark of all of these cases \cite{Zurita, ZuritaResonancias, Sabino, Sabino2018} are band gaps in the wave vector, rather than the frequency. Such gaps were indeed observed for microwaves in a dynamic TL \cite{GenuieGapk, ExtModulado}. We also mention a partial review of temporal photonic crystals and modulated TLs \cite{AlgredoBadilloDr}. All of these papers concern 1D periodicity in space.

Recently, there has been growing interest in spectral singularities in parameter space, these singularities were entitled ``critical points'' \cite{Alu}. The critical points are singularities at which two or more eigenvalues are degenerate, although critical points also exist in other kinds of problems \cite{calculo, Yakovlev}. Two kinds of critical points occur in general in eigenvalue problems that depend on system parameters: exceptional points and ``diabolic points'' \cite{Alu, miller}. The exceptional points are degeneracies where the eigenvalues and corresponding (linearly dependent) eigenvectors coalesce \cite{Alu, Seyranian, heiss}; these points are sensitive to changes in the system parameters and this sensitivity depends on the ``order of the point'' \cite{Chen, Hodei}. These exceptional points are found in optical systems \cite{Alu}, in systems with parity-time (PT) symmetry \cite{ReviewPTRamy}, and also in electronic systems with periodic time variation \cite{Capolino}. On the other hand, diabolic points are accidental degenerations that are found when two branches of eigenvalues with different (linearly independent) eigenvectors cross \cite{Alu, Seyranian, berry}. They are characterized by conical symmetry in parameter space, forming double cones around the point of degeneracy. This structure resembles a south American toy called ``diábolo'' \cite{berry, Alu, miller}; hence the degeneracy point is also known as ``diabolic''. These points are not as sensitive as the exceptional points \cite{Chen} and are well known in quantum systems with ``anti-crossing'' effect \cite{cuantica}. The existence of diabolic points in such systems has been demonstrated experimentally by means of the Berry phase \cite{berry}.

In this paper we explore the properties of EM waves that can propagate in the boundless 2D TL shown in Fig.\ref{Cto_Celda_2D}. While we consider the inductances L to be constant, the capacitances are assumed to be modulated in time periodically, thus $C(t + T) = C(t)$, where $T$ is the modulation period. Thus, we have 2D+1 periodicity, the ``$+1$'' denoting the temporal dimension. Note that all the capacitances are modulated in phase or ``in tandem'' and that resistive elements are neglected. While normally the modulations are accomplished by means of voltage-dependent (non-linear) capacitors or varactors, with all the varactors connected to an external time-periodic voltage source \cite{ExtModulado}, this will not be taken into account in the present work. Our main interest lies in the interplay between 2D spatial periodicity and temporal periodicity of a system.

In Section \ref{Eigen} we employ Kirchhoff's laws to derive an eigenvalue equation for EM waves that can propagate in the ``temporal electric crystal'' of Fig.\ref{Cto_Celda_2D}. This equation is solved in Section \ref{unmodulated} for the static case and in Section \ref{empty} for the limit of vanishing modulation strength. In Section \ref{band_strcuture} the eigenvalue equation is solved numerically for the EM band structures $\omega(k_{x}, k_{y})$. This is done for propagation in the high symmetry [10] and [11] directions in Section \ref{10} and for arbitrary direction of propagation in Section \ref{xy}. We consider three cases of modulation strength: vanishing or ``empty temporal lattice'', ``weak'', and ``strong''. In Section \ref{PD} we explore the diabolic points in the band structures, concluding the article in Section \ref{Conclu}.

%% file: SubFigura1a.tex
\ctikzset{bipoles/length=1.1cm, bipoles/thickness=2.5}

\begin{circuitikz}[scale=0.9]
\draw[color=gray,ultra thin,densely dashed]
(0,0)	to [L] (3,0)
		to [L] (6,0)
		to [L,-.] (9,0)
;
\draw[color=black]
(2,1)	to [L] (5,1)
		to [L] (8,1)
		to [L,-.] (11,1)
;
\draw[color=black]
(4,2)	to [L] (7,2)
		to [L] (10,2)
		to [L,-.] (13,2)
;
\draw[color=gray,densely dashed,ultra thin]
(6,3)	to [L] (9,3)
		to [L] (12,3)
		to [L,-.] (15,3)
;
\draw[color=gray,densely dashed,ultra thin]
(0,0)	to [L] (2,1)
		to [L] (4,2)
		to [L,-.] (6,3)
;
\draw[color=black]
(3,0)	to [L] (5,1)
		to [L] (7,2)
		to [L,-.] (9,3)
;
\draw[color=black]
(6,0)	to [L] (8,1)
		to [L] (10,2)
		to [L,-.] (12,3)
;
\draw[color=gray,ultra thin,densely dashed]
(9,0)	to [L] (11,1)
		to [L] (13,2)
		to [L,-.] (15,3)
;

\draw[color=gray,densely dashed,ultra thin]
(0,-0.6)	to [vC] (0,0)
(2,0.4)	to [vC] (2,1)
(4,1.4)	to [vC] (4,2)
(6,2.4)	to [vC] (6,3)
(9,2.4)	to [vC] (9,3)
(12,2.4)to [vC] (12,3)
(15,2.4)to [vC] (15,3)
(13,1.4)to [vC] (13,2)
(11,0.4)to [vC] (11,1)
(9,-0.6)to [vC] (9,0)
(6,-0.6)to [vC] (6,0)
(3,-0.6)to [vC] (3,0)
;
\draw[color=black,ultra thin,densely dashed]
(0,-0.6) node[tlground]{}
(2,0.4) node[tlground]{}
(4,1.4) node[tlground]{}
(6,2.4) node[tlground]{}
(9,2.4) node[tlground]{}
(12,2.4) node[tlground]{}
(15,2.4) node[tlground]{}
(13,1.4) node[tlground]{}
(11,0.4) node[tlground]{}
(9,-0.6) node[tlground]{}
(6,-0.6) node[tlground]{}
(3,-0.6) node[tlground]{}
;
\draw[color=black]
(5,0.4)	to [vC] (5,1)
(5,0.4)	node[tlground]{}
(7,1.4)	to [vC]	(7,2) 
(7,1.4)	node[tlground]{}
(10,1.4)to [vC]	(10,2)
(10,1.4)node[tlground]{}
(8,0.4)	to [vC] 	(8,1)
(8,0.4)	node[tlground]{}
;

\draw (0,3) node {(a)};
\end{circuitikz}

%% file: SubFigura1b.tex
\ctikzset{bipoles/length=1.2cm, bipoles/thickness=2.8}

\begin{circuitikz}[scale=0.7]

\draw[color=black]
(6,6)	to [L,l=L,f_>=$I^{y}_{p,q}$,-o] (12,6);

\draw[color=black]
(0,6)	to [L,l=L,f>^=$I^{y}_{p,q-1}$,o-] (6,6);

\draw[color=black]
(6,6)	to [L,l_=L,f>=$I^{x}_{p,q}$,-o] (2,4);

\draw[color=black]
(10,8)	to [L,l_=L,f>_=$I^{x}_{p-1,q}$,o-] (6,6);

\draw[color=black]
(6,4)	to [vC,l_=C(t),f<_=$I_{c}$] (6,6);

\draw[color=black]
(2,2.5) node[ground]{}
(0,4.5) node[ground]{}
(10,6.5) node[ground]{}
(12,4.5) node[ground]{}

(6,4) node[ground,rotate=0]{}
;

\draw[color=black]
(0,5.75)		node[anchor=north ] {$V_{p,q-1}$}
(12.5,5.75)	node[anchor=north] {$V_{p,q+1}$}
(2,3.75) 		node[anchor=north ] {$V_{p+1,q}$}
(10,7.65)		node[anchor=north] {$V_{p-1,q}$}
(6.25,6.5) 		node[anchor=east ] {$V_{p,q}$}
;

\draw[color=gray,dotted]
(0,6)	to [L,color=gray,o-] (-4,4)
		to [L,color=gray,-o] (2,4)
(2,4)	to [L,color=gray,o-] (8,4)
		to [L,color=gray,-o] (12,6)		
(12,6)	to [L,color=gray,o-] (16,8)
		to [L,color=gray,-o] (10,8)		
(10,8)	to [L,color=lightgray,o-] (4,8)
		to [L,color=lightgray,-o] (0,6)
;

\draw (-4,8) node {(b)};
\end{circuitikz}

%% file: SecEigenValueProblem.tex
\section{\label{Eigen}Eigenvalue problem for the electromagnetic band structure.}

The object of our investigation is the periodic planar structure of capacitors and inductors shown in Fig.\ref{Cto_Celda_2D}. It is assumed that the capacitors are modulated periodically in time, this external modulation being in tandem, so that all the capacitances in the infinite plane of the structure are given by the same periodic function $C(t)$ at any instant $t$. We note that for constant (unmodulated) capacitances Fig.\ref{Cto_Celda_2D} would simply represent a 2D low-pass transmission line. 

For simplicity, we assume a square lattice in Fig.\ref{Cto2D} and define a unit cell in Fig.\ref{Celda2D}, of side \emph{a}, there is one varactor and four inductors. The origin is placed at an arbitrary vortex and the position of any vortex is specified by the discrete 2D vector $\mathbf{r}=ap\mathbf{\hat{x}}+aq\mathbf{\hat{y}}$, where $p$ and $q$ run over all the integers and $\mathbf{\hat{x}}$ and $\mathbf{\hat{y}}$ are unit vectors in the $x$ and $y$ directions. Thus the pair of integers $(p,q)$ defines an arbitrary unit cell. Applying Kirchhoff's current law to the point $(p,q)$ (see notation in Fig.\ref{Celda2D}) and expressing the voltage over the four adjacent inductors as $LdI^{x,y}_{p,q}/dt$, where $I^{x,y}_{p,q}$ is the appropriate current flowing through the inductor, we obtain the relation
\begin{equation} \label {Ec_cto_2D}
L\frac{d}{dt}I_{C}=V_{p-1,q}+V_{p+1,q}+V_{p,q-1}+V_{p,q+1}-4V_{p,q}.
\end{equation}
Equation (\ref{Ec_cto_2D}) has a plane wave-type spatial solution
\begin{equation} \label {Onda_Plana_2D}
V_{p,q}(\mathbf{r},t)=V(t)e^{i\mathbf{k}\cdot \mathbf{r}}
\end{equation}
where $\mathbf{k}=k_{x}\mathbf{\hat{x}}+k_{y}\mathbf{\hat{y}}$ is the 2D wave vector. Taking into account the dependence on time of the capacitance $C(t)$, the current through it is
\begin{equation} \label {Corriente_Cap_periodica}
I_{C}=\frac{d}{dt}[C(t)V_{p,q}(\mathbf{r},t)].
\end{equation}
Substituting (\ref{Onda_Plana_2D}) and (\ref{Corriente_Cap_periodica}) in (\ref{Ec_cto_2D}), the voltage $V(t)$ is found to satisfy the differential equation
\begin{equation} \label {Ec_cto_2D_1}
\frac{d^{2}}{dt^{2}}[C(t)V(t)]+\frac{4}{L}[sin^{2}(k_{x}a/2)+sin^{2}(k_{y}a/2)]V(t)=0.
\end{equation}
Having assumed that the varactors are modulated periodically in time, $C(t)$ can be expanded in a Fourier series
\begin{equation} \label {Fourier_C}
C(t)=\sum_{n}C_{n}e^{in\Omega t}
\end{equation}
where the (circular) modulation frequency is $\Omega=2\pi/T$, $T$ being the period of modulation, and $n$ runs over all integers. The temporal periodicity also implies that the solutions of (\ref{Ec_cto_2D_1}) must satisfy the Bloch-Floquet theorem. Namely, the potential $V(t)$ must have the form
\begin{equation} \label {BlochFloquet_V}
V(t)=\overline{V}(\omega , t) e^{-i\omega t}
\end{equation}
where the (circular) frequency $\omega$ corresponds to the excitation frequency and $\overline{V}(\omega,t)$ is periodic in time with the period $T=2\pi/\Omega$. Hence, it can be also expanded in a Fourier series
\begin{equation} \label {Fourier_V}
\overline{V}(\omega , t)= \sum_{n}v_{n}(\omega )e^{in\Omega t}.
\end{equation}
At this point it is convenient to normalize the Fourier coefficients of the capacitance $C_{n}$, the excitation frequency $\omega$, and the modulation frequency $\Omega$
\begin{equation} \label {Normalizacion}
\hat{C}_{n}=C_{n} / C_{0}, \qquad \hat{\omega}=\omega / \Omega , \qquad \hat{\Omega}=\Omega / \omega_{0},
\end{equation}
$C_{n}$ is normalized by means of the $n=0$ coefficient $C_{0}$ (equal to the average capacitance), $\omega$ is normalized by the modulation frequency $\Omega$, itself normalized by the resonance frequency of a single unmodulated unit cell ($\omega_{0}=1/\sqrt{LC}$). Then substituting (\ref{Fourier_C})-(\ref{Normalizacion}) into (\ref{Ec_cto_2D_1}) and after some algebra we obtain that
\setlength{\arraycolsep}{0.0em}
\begin{align}
\sum_{n}\Big\{\hat{C}_{m-n} &(\hat{\omega}  -m)^{2} \nonumber\\
{-}\Big( \frac{2}{\hat{\Omega} }\Big)^{2}[sin^{2}(k_{x}a/2)+&sin^{2}(k_{y}a/2)]\delta _{mn}\Big\}v_{n}=0,\nonumber
\end{align}
\setlength{\arraycolsep}{5pt}
\begin{equation} \label {Eq_eigen_2D_normalizada}
m,n=0,\pm 1,\pm 2,\pm 3...
\end{equation}
$\delta_{mn}$ being the Kronecker delta. With the free index $m$ and the summation index $n$ running through all the integer values, (\ref{Eq_eigen_2D_normalizada}) corresponds an infinite number of homogeneous linear equations for the infinite number of unknown Fourier coefficients $v_{n}$ of the potential $V(t)$. Thus, this is an eigenvalue equation; once we set to zero the determinant of the coefficients of $v_{n}$ it is possible to calculate the eigenvalues. In a typical problem the reduced frequency $\hat{\omega}$ and a chosen propagation direction $\mathbf{\hat{k}}$ for the wave are given and the task is to determine the magnitude of the wave vector $k=(\hat{\omega},\mathbf{\hat{k}})$, namely the eigenvalue. That also leads to the wavelength $\lambda=2\pi /k$. Once the eigenvalues have been calculated, the corresponding normalized eigenvectors $v_{n}/v_{0}$ can be found from the full (\ref{Eq_eigen_2D_normalizada}). We note that this equation is periodic in $\hat{\omega}$ with unit period and also periodic in $k_{x}a$ and $k_{y}a$ with period $2\pi$.

Importantly, we have not restricted the form (profile) of modulation. Thus, for example, $C(t)$ could be a harmonic or a square function. We can also consider the special case of static capacitors and the limiting case of vanishing modulation strength. These two related situations are the topics in Section \ref{unmodulated} and Section \ref{empty}.

%% file: SubSecCaseStatic.tex
\subsection{\label{unmodulated}The special case of static (unmodulated) capacitors.}

The absence of modulation is equivalent to the assumption of vanishing modulation frequency $\Omega=0$. With this substitution in (\ref{Eq_eigen_2D_normalizada}) we get that
\begin{equation} \label {RelDispersion_estatico}
\omega = 2 \omega_{0}[sin^{2}(k_{x}a/2)+sin^{2}(k_{y}a/2)]^{1/2}.
\end{equation}
This is the dispersion relation for the 2D low-pass transmission line \cite{lbrillouin} \cite{RelDisp_Estatica}. It corresponds to propagation up to the frequency $\omega=2\omega_{0}$ in a [10] direction ($k_{y}=0$) and up to $\omega=2\sqrt{2}\omega_{0}$ in a [11] direction ($k_{x}=k_{y}$).

%% file: SubSecCaseEmptyRed.tex
\subsection{\label{empty}The limit of vanishing modulation strength (empty temporal lattice).}

Unlike the case \ref{unmodulated}, here we wish to preserve the modulation, namely $\Omega \neq 0$, albeit with vanishing strength. This implies that in (\ref{Fourier_C}) all the Fourier coefficients of the capacitance are taken to vanish, with the exception of $C_{n=0}$. Then $C(t)=C_{0}$ and in (\ref{Normalizacion}) $\hat{C}_{n}=\delta_{n0}$. Substituting $\hat{C}_{m-n}=\delta_{mn}$ in (\ref{Eq_eigen_2D_normalizada}) we find that
\begin{equation} \label {Eq_eigen_2D_empty_simplify}
\hat{\omega} =\pm \Big( \frac{2}{\hat{\Omega} }\Big)[sin^{2}(k_{x}a/2)+sin^{2}(k_{y}a/2)]^{1/2}+m.
\end{equation}
This is the dispersion relation for the 2D empty temporal lattice. Once we replace $\hat{\omega}$ by $\omega/\Omega$,  $\hat{\Omega}$ by $\Omega/\omega_{0}$ (see (\ref{Normalizacion})), $m=0$, and take the positive root, (\ref{Eq_eigen_2D_empty_simplify}) gives a solution that is identical to the static dispersion relation (\ref{RelDispersion_estatico}). Hence, we realize that the modulation (even when having vanishing strength) produces two qualitative changes: one is that the low frequency band given by (\ref{RelDispersion_estatico}) now also has an inversion, as given by the negative sign, the second difference is that both of these bands also become shifted by positive (negative) integers $m$ upward (downward) in frequency.

The upward ($+$) $m=0$ band and the downward ($-$) $m=1$ band are related in an important way. It is not difficult to verify that they are reflections of each other about the line $\hat{\omega}=1/2$. That is, $\hat{\omega}^{+}_{0}=1-\hat{\omega}_{1}^{-}$ where we introduced the notation $\hat{\omega}_{m}^{\pm }$ for the frequency eigenvalues. The width of all 2D allowed bands (determined by the condition $k_{x}a=k_{y}a=\pi$ in (\ref{Eq_eigen_2D_empty_simplify})) is $2\sqrt{2}/\hat{\Omega}$. If $\hat{\Omega}=4\sqrt{2}$ the extrema of these $\pm $ bands will just touch at $\hat{\omega}=1/2$. On the other hand, for $\hat{\Omega}>4\sqrt{2}$ there is a frequency band gap (centered at $\hat{\omega}=1/2$), while for $\hat{\Omega}<4\sqrt{2}$ the upward and downward bands overlap, as seen in Fig.\ref{AtlasVacia} (for the [11] direction). Thus, the behavior changes qualitatively, depending whether the modulation frequency $\Omega$ is equal to, greater than or smaller than $4\sqrt{2}$ times the resonances frequency $\omega_{0}$.

The corresponding limit in solid state physics (with crystalline potentials $V(\vec{r})\rightarrow constant$) is known as the empty lattice model \cite{Kittel}. In the case of modulation in time we call it the empty temporal lattice model \cite{Sabino2018}.

%% file: SecBandsStructure.tex
\section{\label{band_strcuture}Electromagnetic band structures.}

This section is devoted to solutions of the eigenvalue problem $\hat{\omega}=\hat{\omega}(k_{x},k_{y})$ based on the eigenvalue equation (\ref{Eq_eigen_2D_normalizada}) for the normalized variables.
We assume that the capacitances are modulated harmonically
\begin{equation} \label {Cap_temporal}
C(t)=C_{0} [1+m_{c} sin(\Omega t)]
\end{equation}
here $m_{c}$ is the strength of modulation or, briefly, just ``modulation''. Because $C(t)$ can not be negative, $m_{c}$ ranges from 0 to 1. Throughout our numerical work it will take three values: $m_{c}=0$ for the empty temporal lattice, $m_{c}=0.36$ (a relatively weak modulation), and $m_{c}=0.86$ (a relatively strong modulation). In order to made more manageable our complex 2D+1 band structure problem, first we present results for propagation in a [10] (with $k_{y}=0$) and [11] (with $k_{x}=k_{y}$) directions and then to address propagation in an arbitrary direction $\mathbf{k}$ in the plane in Section \ref{xy}.

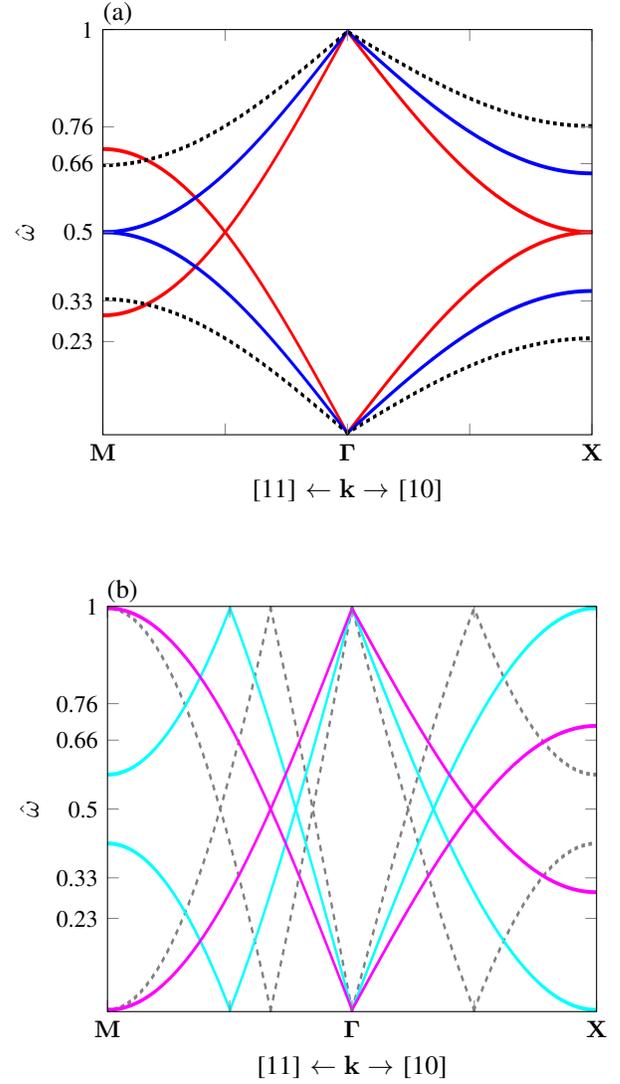
\begin{figure}[!t]
\captionsetup[subfigure]{labelformat=empty}
\centering
\subfloat[\label{subfig:2a}]{\input{SubFigura2a}}
\hspace{-0.3cm}
\subfloat[\label{subfig:2b}]{\input{SubFigura2b}}
\caption{Band structures $\hat{\omega}$ versus $k$ in the limit of vanishing modulation, $m_{c} \rightarrow 0$ (empty temporal lattice). The two high symmetry directions of propagation $\mathbf{\Gamma }$ $\mathbf{X}$ (or [10]) and $\mathbf{\Gamma }$ $\mathbf{M}$ (or [11]) are considered. In (a) the dotted black curve, blue curve, and red curve are characterized, respectively, by the parameters $\hat{\Omega}(=\Omega/\omega_{0})=6\sqrt{2}$, $4\sqrt{2}$, and $4$. In (b) the magenta, cyan, and gray curves correspond to $\hat{\Omega}=2\sqrt{2}$, $2$, and $\sqrt{2}$.}
\label{AtlasVacia}
\end{figure}

%% file: SubFigura2a.tex
\pgfplotsset{width=230.0pt}

\begin{tikzpicture}
\begin{axis}[
enlargelimits=false,
axis on top,
ylabel={$\hat{\omega}$},
ytick={0.23,0.33,0.5,0.669,0.76,1},
xlabel={[11] $\leftarrow\mathbf{k}\rightarrow $ [10]},
xtick={-3.1416,-1.5708,0,1.5708,3.1416},
xticklabels={$\mathbf{M}$, ,$\mathbf{\Gamma}$, ,$\mathbf{X}$},
yticklabels={0.23,0.33,0.5,0.66,0.76,1},
]
\addplot graphics [
xmin=-3.1416 ,xmax=3.1416 ,ymin=0,ymax=1,
includegraphics={trim=6.9 7.2 6.9 6.15,clip},
] {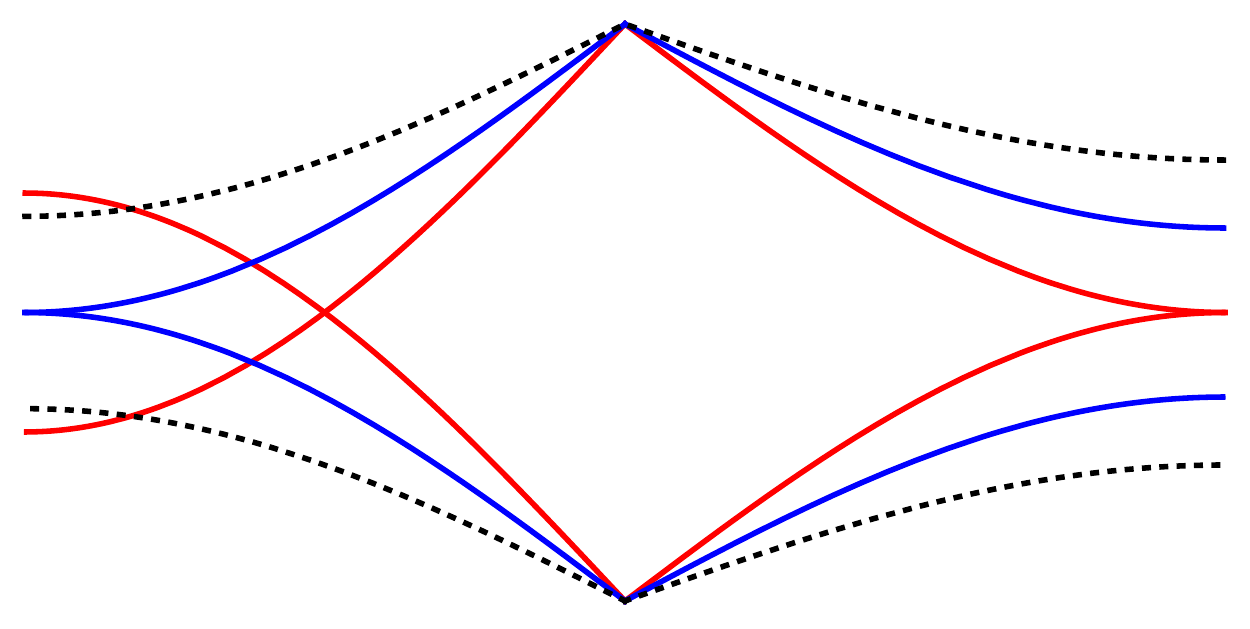};
\end{axis}
\draw (0.2,5.6) node {(a)};
\end{tikzpicture}

%% file: SubFigura2b.tex
\pgfplotsset{width=230.0pt}

\begin{tikzpicture}
\begin{axis}[
enlargelimits=false,
axis on top,
ylabel={$\hat{\omega}$},
ytick={0.23,0.33,0.5,0.669,0.76,1},
xlabel={[11] $\leftarrow\mathbf{k}\rightarrow $ [10]},
ytick={0.23,0.33,0.5,0.67,0.76,1},
xtick={-3.1416,-1.5708,0,1.5708,3.1416},
xticklabels={$\mathbf{M}$, ,$\mathbf{\Gamma}$, ,$\mathbf{X}$},
yticklabels={0.23,0.33,0.5,0.66,0.76,1},
]
\addplot graphics [
xmin=-3.1416 ,xmax=3.1416 ,ymin=0,ymax=1,
includegraphics={trim=6.9 7.2 6.9 6.15,clip},
] {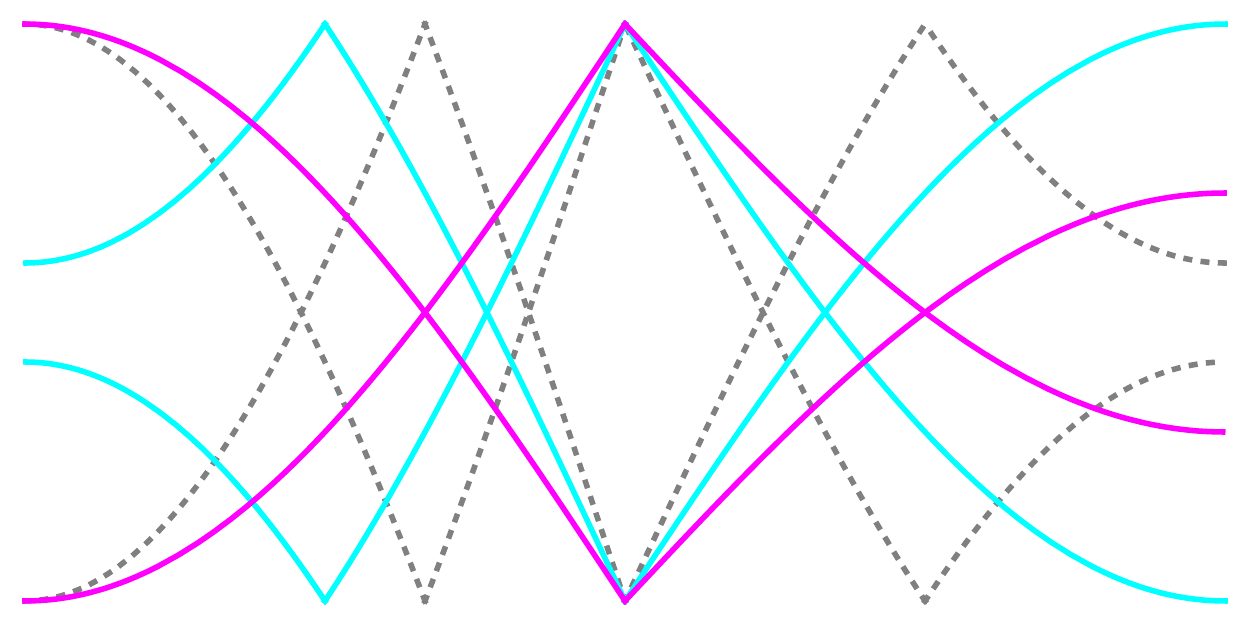};
\end{axis}
\draw (0.2,5.6) node {(b)};
\end{tikzpicture}%

%% file: SubSecDir10_11.tex
\subsection{\label{10}Propagation in the \hkl[10] and \hkl[11] directions.}

\begin{figure*}[!h] 
\captionsetup[subfigure]{labelformat=empty,nearskip=-1cm,farskip=-1cm}
\centering
\subfloat[\label{8atls1939}]{}%
\subfloat[\label{7atls1658}]{}%
\subfloat[\label{6atls1240}]{}%
\subfloat[\label{5atls1430}]{}%
\subfloat[\label{4atls5201}]{}%
\subfloat[\label{4atls0002}]{}%
\subfloat[\label{3atls4414}]{}%
\subfloat[\label{2atls4496}]{}%
\subfloat[\label{1atls8849}]{}%
\input{Figura3}%
\caption{Atlas of band structures for propagation in the [10] (or $\mathbf{\Gamma }$ $\mathbf{X}$) direction; two spatial BZs and a single temporal BZ is presented. Each subfigure is characterized by some value of the parameter $\hat{\Omega}$, in descending order from $\hat{\Omega}=8.1939$ in (a) to $\hat{\Omega}=1.8849$ in (i). The modulation parameter $m_{c}$ takes the values 0 (empty temporal lattice), 0.36 (weak modulation), and 0.86 (strong modulation) for the dotted black lines, red lines, and blue lines, respectively. The horizontal yellow bands indicate forbidden frequency bands, light yellow for $m_{c} = 0.86$, butter yellow for $m_{c} = 0.36$, and deep yellow for $m_{c} \rightarrow 0$. The vertical green bands indicate forbidden wave vector bands, strong green for $m_{c} = 0.86$, and lemon green for $m_{c} = 0.36$.}
\label{Atlas01}
\end{figure*}
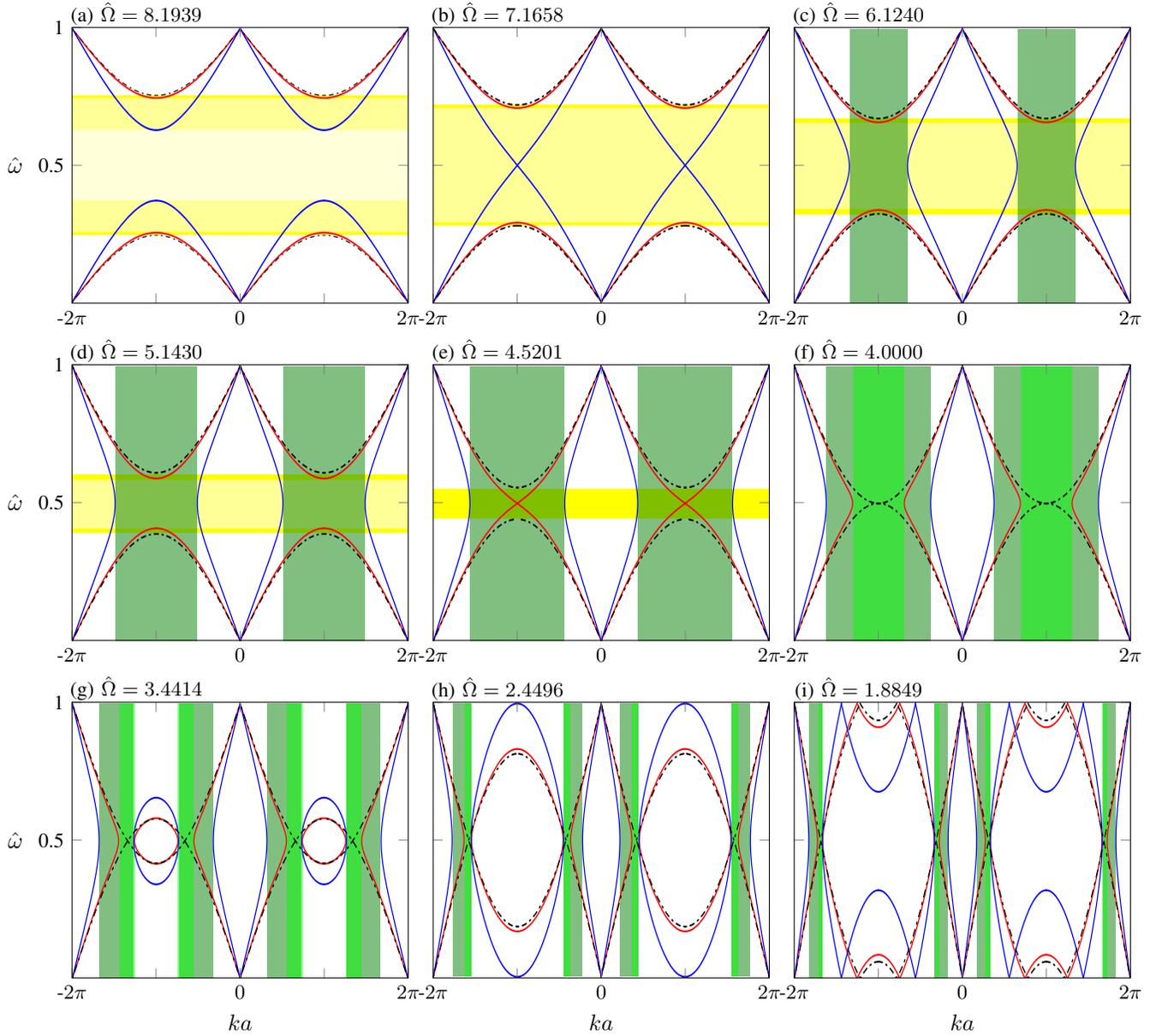

For propagation in the [10] direction, all the voltages along any line of varactors that is perpendicular to the $x$ axis have the same voltage (in magnitude and phase), for which reason there is no current in the $y$ direction. Because in our simple model we have excluded resistive elements, these plane waves propagate without attenuation.

Consider first the unmodulated TL ($\Omega=0$). By (\ref{RelDispersion_estatico}) the dispersion relation is simply
\begin{equation} \label {RelDispersion_estatico_10}
\omega /\omega_{0} = 2|sin(k_{x}a/2)|
\end{equation}
this is identical to the dispersion relation for vibrational waves in a mono-atomic lattice \cite{Kittel}. For very low frequencies, $\omega \ll \omega_{0}$, (\ref{RelDispersion_estatico_10}) is identical to the dispersion relation for the ordinary 1D TL and gives the well known phase (and group) velocity $\omega/k_{x}=1/\sqrt{\bar{L}\bar{C}_{0}}$, where $\bar{L}$ and $\bar{C}_{0}$ are the distributed (per unit length) inductance and capacitance \cite{Pozar}. With increasing frequency the group velocity $d\omega/dk_{x}$ decreases and vanishes at the edge of the BZ $k_{x}=\pi/a$. For this value of $k_{x}$ the band reaches its upper edge at $\omega=2\omega_{0}$.

Now we turn to the case of vanishing modulation strength, $m_{c}=0$ in (\ref{Cap_temporal}), albeit $\Omega\neq 0$. And $k_{y}=0$ in (\ref{Eq_eigen_2D_empty_simplify}), gives
\begin{equation} \label {Eq_eigen_1D_empty_simplify}
\hat{\omega} =\pm \Big( \frac{2}{\hat{\Omega} }\Big)[sin(k_{x}a/2)]+m
\end{equation}
\[m=0,\pm 1, \pm 2...\]
for this empty temporal lattice model. Taking into account the two adjacent temporal BZs, the width of the allowed bands is $4/\hat{\Omega}$. Thus for very large $\hat{\Omega}$ (such that $\Omega\gg 2\omega_{0}$) the bands shrink to negligible width, while in the opposite limit the bands overlap so that no frequency is prohibited for propagation. It is instructive to single out the $\hat{\omega}^{+}_{0}$ and $\hat{\omega}^{-}_{1}$ bands. These bands are reflections of each other about the line $\hat{\omega}=1/2$, as seen in Fig.\ref{AtlasVacia} for  series of values of $\hat{\Omega}$, as we mentioned before. If $\hat{\Omega}=4$ the $\hat{\omega}^{+}_{0}$ and $\hat{\omega}^{-}_{1}$ bands just touch in [10] direction. For a band gap to form it is necessary to have a band width smaller than $1/2$, namely $\hat{\Omega}>4$. On the other hand, for $\hat{\Omega}<4$ the upward and downward bands overlap. Because of the unit periodicity of $\hat{\omega}$, this behavior is repeated in all the temporal BZs, one of which is depicted in Fig.\ref{AtlasVacia}. Note that, for any value of $\hat{\Omega}$ the $\hat{\omega}^{+}_{0}$ bands dispersion is given by (\ref{RelDispersion_estatico_10}) for the unmodulated TL. For the [11] direction, the band structure has the same qualitative behavior, however, the aforementioned values of $\hat{\Omega}$  are multiplied by $\sqrt{2}$, as seen in Fig.\ref{AtlasVacia}.

In order to obtain the spatio-temporal band structure for finite modulation strength, we solved the eigenvalue equation (\ref{Eq_eigen_2D_normalizada}) (with $k_{y}=0$). The results are pictured in Fig.\ref{Atlas01} for two spatial BZs ($-2\pi\leq k_{x}a \leq 2\pi$) and one temporal BZ ($0 \leq \hat{\omega} \leq 1$). The modulation parameter $m_{c}$ is represented by three values: $m_{c}=0$ (empty temporal lattice) dotted black line, $m_{c}=0.36$ (weak modulation) red line, and $m_{c}=0.86$ (strong modulation) blue line. The other parameter, namely the reduced modulation frequency $\hat{\Omega}$ takes on nine values, starting with $\hat{\Omega}=8.1939$ in Fig.\ref{8atls1939} and descending in unequal steps to $\hat{\Omega}=1.8849$ in Fig.\ref{1atls8849}. The periodicity in $k_{x}a$ (with period $2\pi$) and the reflection symmetry about $\hat{\omega}=1/2$ are evident. Although not shown, the same band structure is to be repeated periodically into other temporal BZs such as $1 \leq \hat{\omega} \leq 2$.

Generally speaking, in Fig.\ref{Atlas01}, the band structures for $m_{c}=0.36$ and $m_{c}=0$ are very similar, so indeed, modulations $m_{c} \leq 0.36$ can be considered ``weak''. Nevertheless, qualitative differences can be observed in some cases in the region $\hat{\omega}\sim 1/2$, especially in Fig.\ref{4atls5201} and Fig.\ref{4atls0002}. On the other hand, the band structures for $m_{c}=0.86$ differ in an important way for all $\hat{\Omega}$ from those for $m_{c}=0.36$, thus justifying the term ``strong modulation'' for $m_{c}=0.86$. We confirm the behavior found in Fig.\ref{AtlasVacia} for $m_{c}=0$, that a frequency gap exists for $\hat{\Omega} > 4$ increasing in width with $\hat{\Omega}$ and there is no \emph{wave vector} gap for any value of $\hat{\Omega}$. 

In the case of weak modulation (red lines), we find an $\omega$-gap for $\hat{\Omega}>4.5201$, see Fig.\ref{8atls1939}-\ref{5atls1430}. The gap closes for $\hat{\Omega}=4.5201$, Fig.\ref{4atls5201} (compared with $\hat{\Omega}=4$ in the case of $m_{c}=0$). For $\hat{\Omega}<4.5201$, see Fig.\ref{4atls0002}-\ref{1atls8849}, there are no more frequency gaps; on the other hand, there develop wave vector gaps. For $\hat{\Omega}=4$ in Fig.\ref{4atls0002} these gaps border the BZ edges $k=\pm \pi/a$, however move away from the edge and become narrower as $\hat{\Omega}$ decreases further, Fig.\ref{3atls4414}-\ref{1atls8849}. The maximum of the $+$ band steadily increases as $\hat{\Omega}$ decreases, eventually penetrating into the second temporal BZ, see Fig.\ref{1atls8849}. An identical lobe penetrates from below into the first BZ; corresponding symmetrical behavior can be also observed for the - bands (``+ band'' that curves upward and a ``- band'' curving downward).

In the case of the strong modulation (blue lines) frequency gaps exist only for $\hat{\Omega} > 7.1658$, although much narrower than for weak modulation, see Fig.\ref{8atls1939}. For $\hat{\Omega} = 7.1658$ the gap closes, as observed in Fig.\ref{7atls1658} and for $\hat{\Omega} < 7.1658$ wave vector gaps develop, Fig.\ref{6atls1240}-\ref{1atls8849}. These are considerably wider than for weak modulation. For $\hat{\Omega}=3.4414$ (Fig.\ref{3atls4414}) a separate structure develops at the spatial BZ edges and, as $\hat{\Omega}$ is decreased, the extrema of this structure move further apart, until for $\hat{\Omega}=2.4496$ in Fig.\ref{2atls4496} they just touch the boundaries of the temporal BZ at $\hat{\omega}=1$ and $\hat{\omega}=0$. For still smaller $\hat{\Omega}$ penetration of lobes from neighboring temporal BZs can be seem, as noted for the weak modulation, but now much stronger.

We call attention to the Fig.\ref{4atls5201} and Fig.\ref{7atls1658} where for the modulation 0.36 and 0.86, respectively, neither a $\omega$-gap nor a $k$-gap develops. The corresponding values of the modulation frequency $\hat{\Omega}$, 4.5201 and 7.1658 are transition points between $\omega$-gaps for greater values of $\hat{\Omega}$ and $k$-gaps for smaller values of $\hat{\Omega}$. Sec. \ref{PD} will be devoted to such ``phase transitions''.

For the [11] direction, again, the qualitative behavior is the same, however with the special values of $\hat{\Omega}$ multiplied by $\sqrt{2}$.

%% file: Figura3.tex
\pgfplotsset{width=0.375\linewidth}

\hspace{-0.25cm}%
\begin{tikzpicture}
\begin{axis}[
enlargelimits=false,
axis on top,
ylabel={$\hat{\omega}$},
every axis y label/.style={
at={(ticklabel cs:0.5)},
anchor=near ticklabel,
rotate=false,
},
ytick={0.5,1},
xtick={-6.2832,-3.1416,0,3.1416,6.2832},
xticklabels={-$2\pi$, ,$0$, ,$2\pi$},
yticklabels={0.5 ,1},
]
\addplot graphics [
xmin=-6.2832 ,xmax=6.2832 ,ymin=0,ymax=1,
includegraphics={trim=6.9 7.2 6.9 6.15,clip},
] {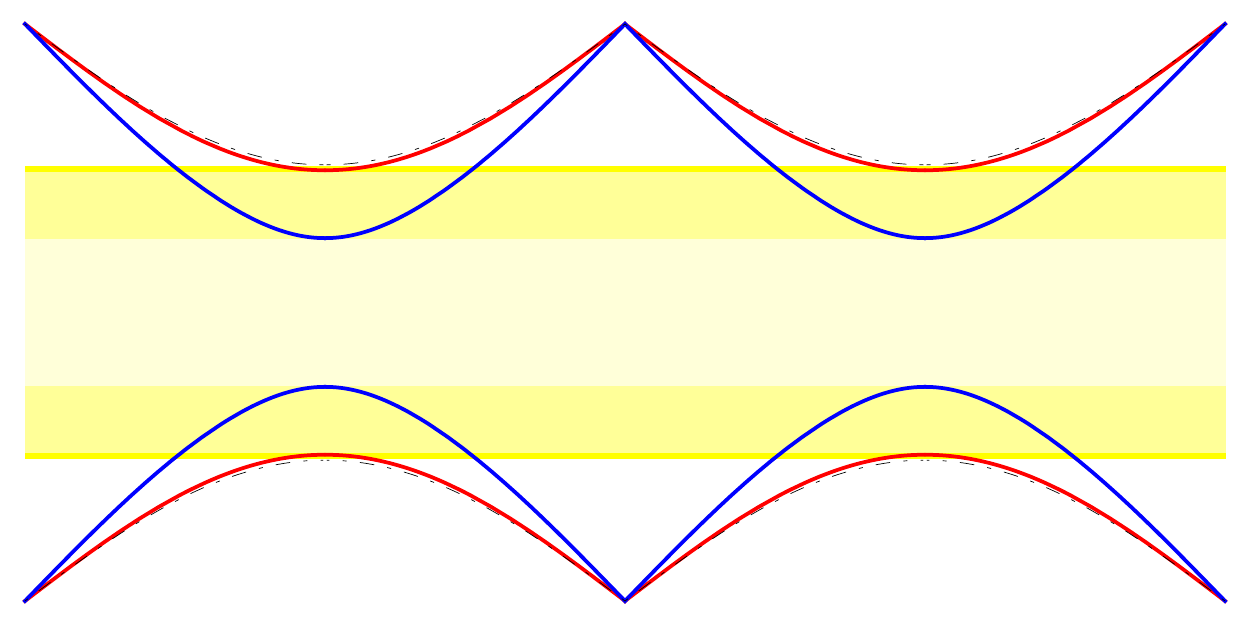};
\end{axis}
\draw (1.0,4.5) node {\small (a) $\hat{\Omega} =8.1939$};
\end{tikzpicture}%
\hspace{-0.26cm}%
\begin{tikzpicture}
\begin{axis}[
enlargelimits=false,
axis on top,
ytick={0.5,1},
xtick={-6.2832,-3.1416,0,3.1416,6.2832},
xticklabels={-$2\pi$, ,$0$, ,$2\pi$},
yticklabels={,},
]
\addplot graphics [
xmin=-6.2832 ,xmax=6.2832 ,ymin=0,ymax=1,
includegraphics={trim=6.9 7.2 6.9 6.15,clip},
] {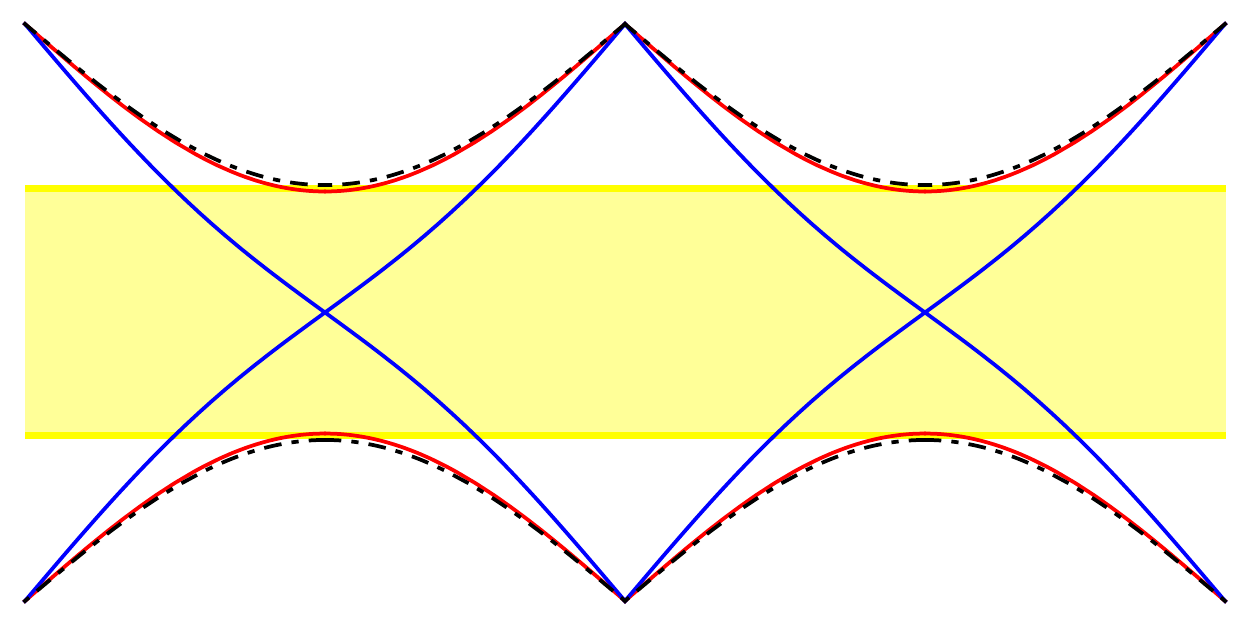};
\end{axis}
\draw (1.0,4.5) node {\small (b) $\hat{\Omega} =7.1658$};
\end{tikzpicture}%
\hspace{-0.26cm}%
\begin{tikzpicture}
\begin{axis}[
enlargelimits=false,
axis on top,
ytick={0.5,1},
xtick={-6.2832,-3.1416,0,3.1416,6.2832},
xticklabels={-$2\pi$, ,$0$, ,$2\pi$},
yticklabels={,},
]
\addplot graphics [
xmin=-6.2832 ,xmax=6.2832 ,ymin=0,ymax=1,
includegraphics={trim=6.9 7.2 6.9 6.15,clip},
] {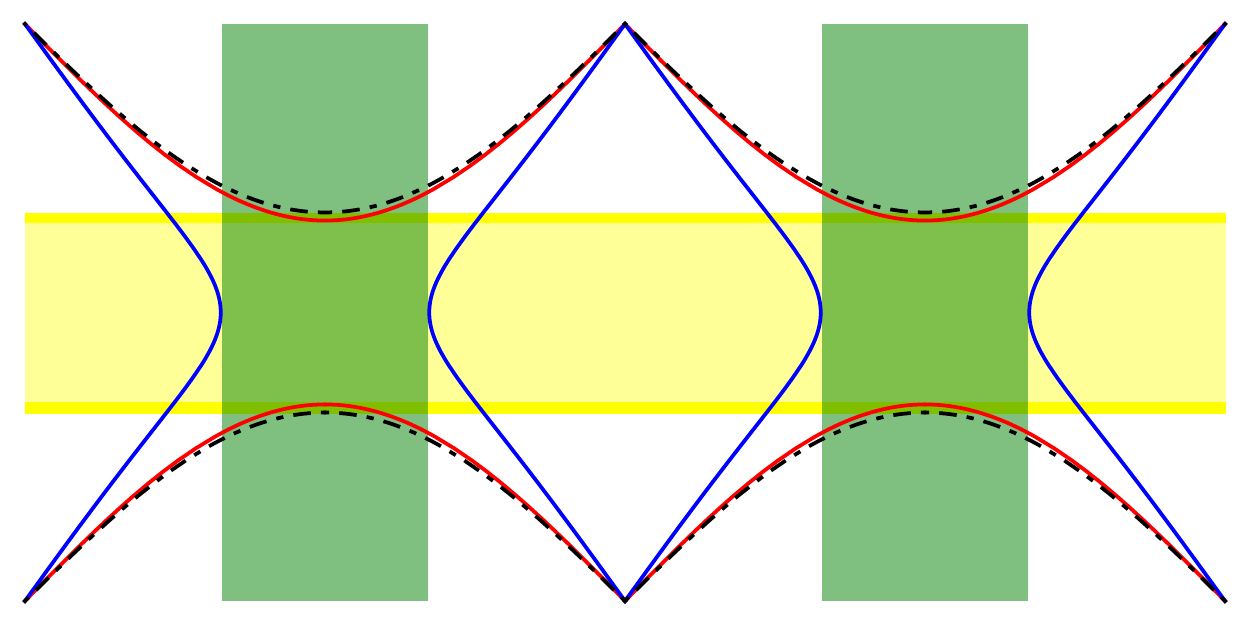};
\end{axis}
\draw (1.0,4.5) node {\small (c) $\hat{\Omega} =6.1240$};
\end{tikzpicture}%

\hspace{-0.25cm}%
\begin{tikzpicture}
\begin{axis}[
enlargelimits=false,
axis on top,
ylabel={$\hat{\omega}$},
every axis y label/.style={
at={(ticklabel cs:0.5)},
anchor=near ticklabel,
rotate=false,
},
ytick={0.5,1},
xtick={-6.2832,-3.1416,0,3.1416,6.2832},
xticklabels={-$2\pi$, ,$0$, ,$2\pi$},
yticklabels={0.5 ,1},
]
\addplot graphics [
xmin=-6.2832 ,xmax=6.2832 ,ymin=0,ymax=1,
includegraphics={trim=6.9 7.2 6.9 6.15,clip},
] {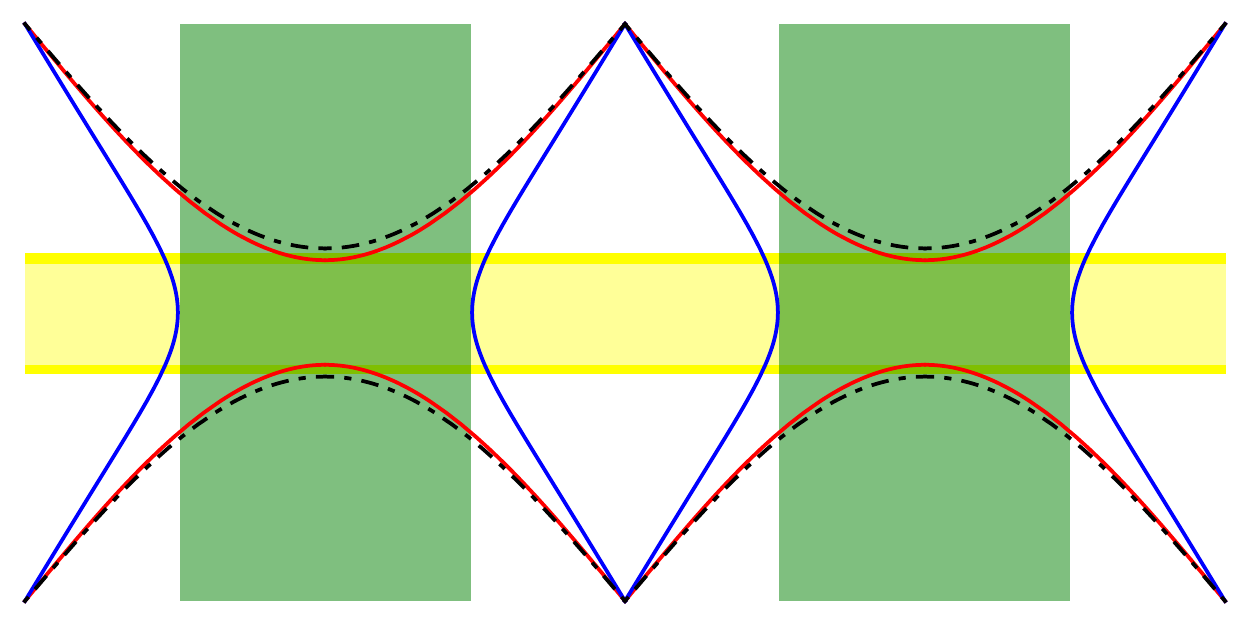};
\end{axis}
\draw (1.0,4.5) node {\small (d) $\hat{\Omega} =5.1430$};
\end{tikzpicture}%
\hspace{-0.26cm}%
\begin{tikzpicture}
\begin{axis}[
enlargelimits=false,
axis on top,
ytick={0.5,1},
xtick={-6.2832,-3.1416,0,3.1416,6.2832},
xticklabels={-$2\pi$, ,$0$, ,$2\pi$},
yticklabels={,},
]
\addplot graphics [
xmin=-6.2832 ,xmax=6.2832 ,ymin=0,ymax=1,
includegraphics={trim=6.9 7.2 6.9 6.15,clip},
] {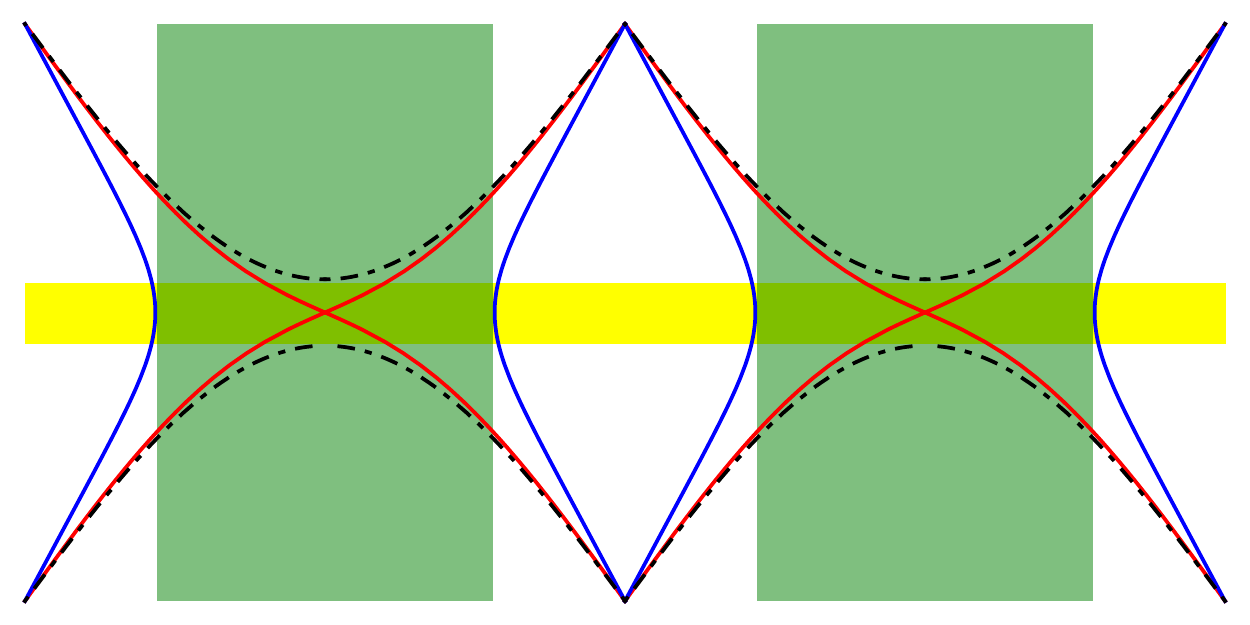};
\end{axis}
\draw (1.0,4.5) node {\small (e) $\hat{\Omega} =4.5201$};
\end{tikzpicture}%
\hspace{-0.26cm}%
\begin{tikzpicture}
\begin{axis}[
enlargelimits=false,
axis on top,
ytick={0.5,1},
xtick={-6.2832,-3.1416,0,3.1416,6.2832},
xticklabels={-$2\pi$, ,$0$, ,$2\pi$},
yticklabels={,},
]
\addplot graphics [
xmin=-6.2832 ,xmax=6.2832 ,ymin=0,ymax=1,
includegraphics={trim=6.9 7.2 6.9 6.15,clip},
] {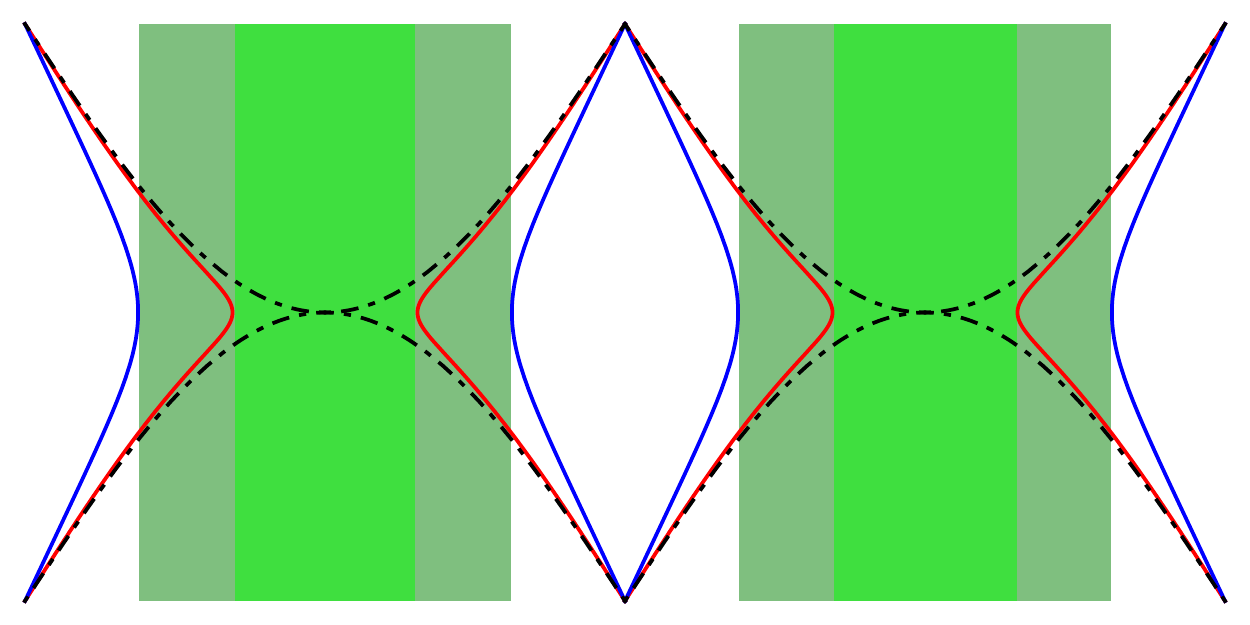};
\end{axis}
\draw (1.0,4.5) node {\small (f) $\hat{\Omega} =4.0000$};
\end{tikzpicture}%

\hspace{-0.25cm}%
\begin{tikzpicture}
\begin{axis}[
enlargelimits=false,
axis on top,
xlabel={$ka$},
ylabel={$\hat{\omega}$},
every axis y label/.style={
at={(ticklabel cs:0.5)},
anchor=near ticklabel,
rotate=false,
},
ytick={0.5,1},
xtick={-6.2832,-3.1416,0,3.1416,6.2832},
xticklabels={-$2\pi$, ,$0$, ,$2\pi$},
yticklabels={0.5 ,1},
]
\addplot graphics [
xmin=-6.2832 ,xmax=6.2832 ,ymin=0,ymax=1,
includegraphics={trim=6.9 7.2 6.9 6.15,clip},
] {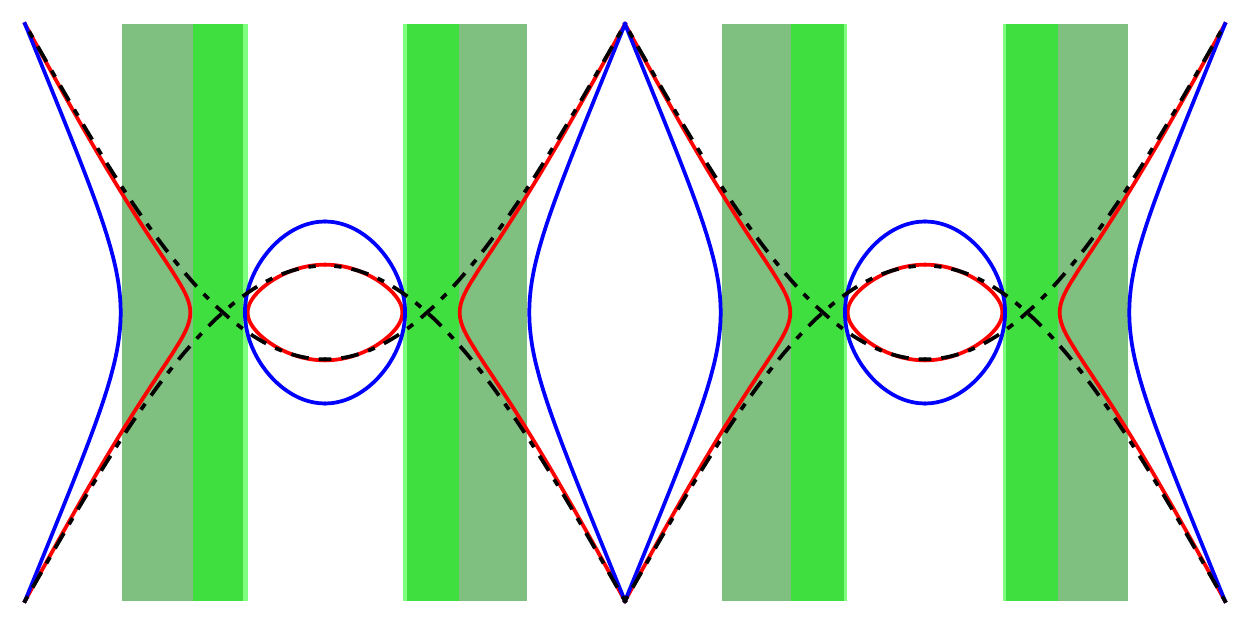};
\end{axis}
\draw (1.0,4.5) node {\small (g) $\hat{\Omega} =3.4414$};
\end{tikzpicture}%
\hspace{-0.26cm}%
\begin{tikzpicture}
\begin{axis}[
enlargelimits=false,
axis on top,
xlabel={$ka$},
ytick={0.5,1},
xtick={-6.2832,-3.1416,0,3.1416,6.2832},
xticklabels={-$2\pi$, ,$0$, ,$2\pi$},
yticklabels={,},
]
\addplot graphics [
xmin=-6.2832 ,xmax=6.2832 ,ymin=0,ymax=1,
includegraphics={trim=6.9 7.2 6.9 6.15,clip},
] {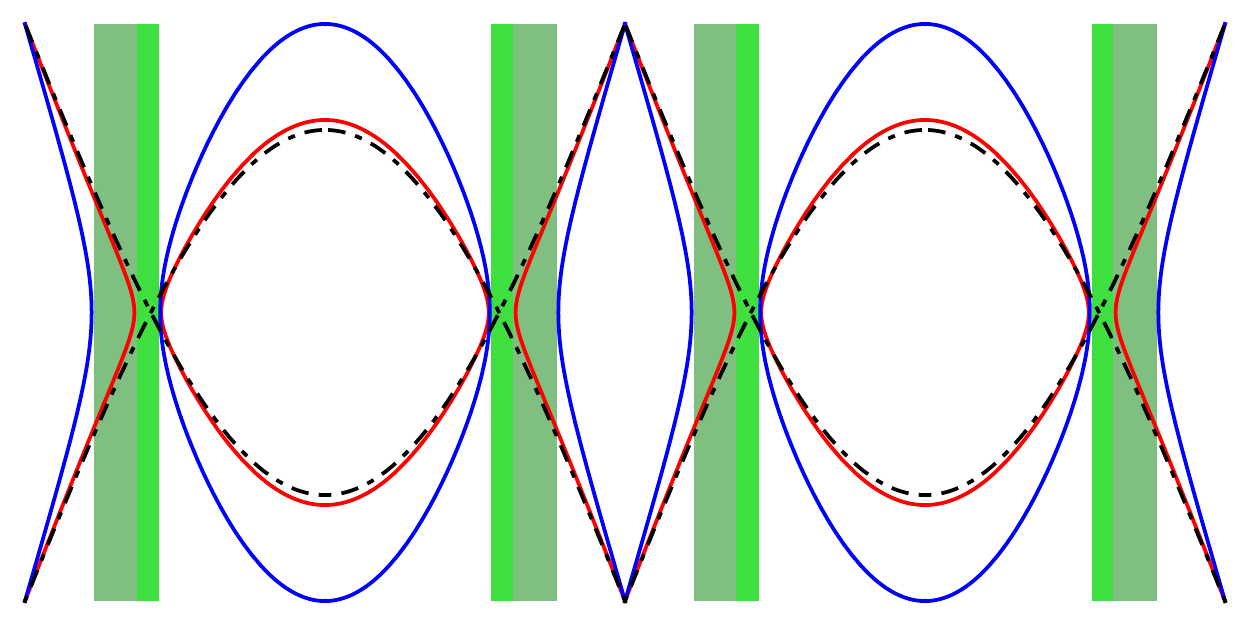};
\end{axis}
\draw (1.0,4.5) node {\small (h) $\hat{\Omega} =2.4496$};
\end{tikzpicture}%
\hspace{-0.26cm}%
\begin{tikzpicture}
\begin{axis}[
enlargelimits=false,
axis on top,
xlabel={$ka$},
ytick={0.5,1},
xtick={-6.2832,-3.1416,0,3.1416,6.2832},
xticklabels={-$2\pi$, ,$0$, ,$2\pi$},
yticklabels={,},
]
\addplot graphics [
xmin=-6.2832 ,xmax=6.2832 ,ymin=0,ymax=1,
includegraphics={trim=6.9 7.2 6.9 6.15,clip},
] {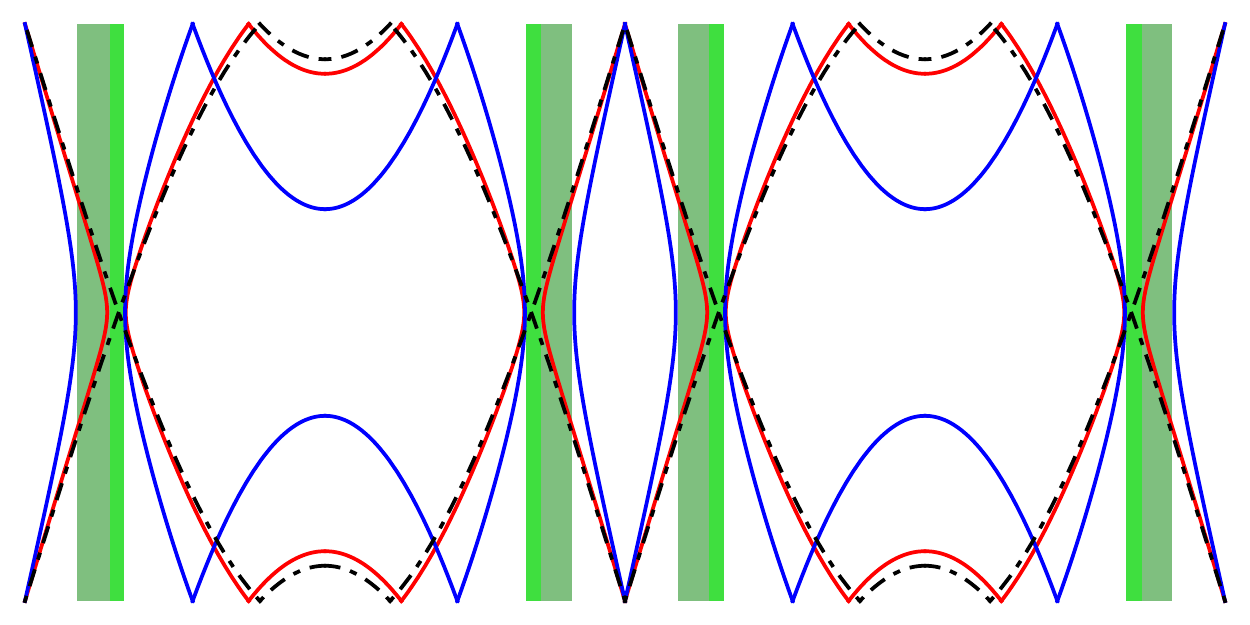};
\end{axis}
\draw (1.0,4.5) node {\small (i) $\hat{\Omega} =1.8849$};
\end{tikzpicture}

%% file: SubSecArbitraryDir.tex
\subsection{\label{xy}Propagation in an arbitrary direction $\mathbf{k}$ in the $x-y$ plane.}

In this subsection we generalize the former Section \ref{10} to propagation in any direction $\mathbf{k}$ in the plane of the 2D TL, not necessarily of high symmetry. First consider the case of static capacitors ($C(t)=C_{0}$), namely a 2D low pass TL. The dispersion is given by (\ref{RelDispersion_estatico}) and we present it as both, isofrequency level contours (like in \cite{lbrillouin}) $\omega/\omega_{0}=const$ in Fig.\ref{Contornos_2D} and a three dimensional surface $\omega/\omega_{0}=\omega(k_{x}a,k_{y}a)/\omega_{0}$ in Fig.\ref{RD_Estatico2D}. The normalized frequency $\omega/\omega_{0}$ varies from $0$ to $2\sqrt{2}$, these limiting values corresponding to the band edges. In Fig.\ref{Contornos_2D}, subsequent contours vary by $\delta\omega/\omega_{0}=0.2$ steps. For $\omega/\omega_{0}	\leq 1$ the contours are nearly circular, however as the (square) BZ edges are approached the anisotropy becomes apparent, especially upon comparison of the \hkl<10> and \hkl<11> directions. As we have noted previously, for the \hkl<10> direction propagation is limited to frequencies below $\omega/\omega_{0}=2$; at this point the isofrequency contour is a square. For $\omega/\omega_{0}>2$ the curvature changes sign and for $\omega/\omega_{0} \approx  2\sqrt{2}$ the isofrequency contours become circles centered at the BZ edges $(k_{x}a,k_{y}a)=(\pm \pi,\pm \pi)$. 

\begin{figure}[!t]
\captionsetup[subfigure]{labelformat=empty}
\centering
\subfloat[\label{Contornos_2D}]{\input{SubFigura4a}}

\subfloat[\label{RD_Estatico2D}]{\input{SubFigura4b}}%
\caption{Dispersion $\omega(k_{x}a, k_{y}a)$ for 2D spatial periodicity with no temporal modulation $(\Omega = 0)$. (a) 14 isofrequency contours from $\omega/\omega_{0} = 0.2$ to $\omega/\omega_{0} = 2.8$ with increments of $0.2$. (b) Three-dimensional rendition of the same. The irreducible BZ $\mathbf{\Gamma }$ $\mathbf{X}$ $\mathbf{M}$ is shown and the high symmetry directions $\mathbf{\Gamma }$ $\mathbf{X}$ ($\mathbf{\Gamma }$ $\mathbf{M}$) are marked in red (blue).}
\label{Estatico_2D}
\end{figure}
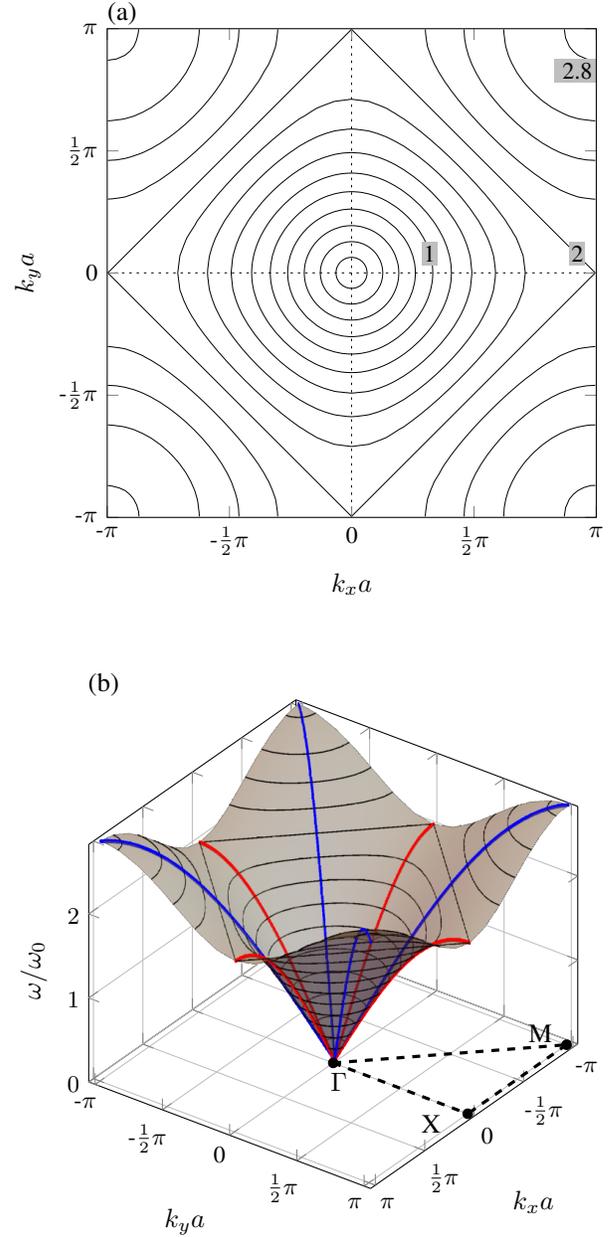

\begin{figure*}[!t] 
\captionsetup[subfigure]{labelformat=empty,nearskip=-1cm,farskip=-1cm}
\centering
\subfloat[\label{8temp1939}]{}%
\subfloat[\label{6temp3926}]{}%
\subfloat[\label{5temp6198}]{}%
\subfloat[\label{5temp0656}]{}%
\subfloat[\label{4temp5201}]{}%
\subfloat[\label{4temp0232}]{}%
\subfloat[\label{3temp6636}]{}%
\subfloat[\label{2temp9133}]{}%
\subfloat[\label{1temp9869}]{}%
\input{Figura5}
\caption{Atlas of band structures for the 2D temporal electric crystal, for 9 values of the parameter $\hat{\Omega}$. Weak modulation $(m_{c}=0.36)$ is considered, presenting a single spatial BZ and a single temporal BZ. In (a) the frequency gap is projected in gray at the surfaces $k_{x}a =\pi$ and $k_{y}a= \pi$. In (c)-(f) the wave vector gaps are projected in gray at the $\hat{\omega}=0$ surface.}
\label{Temporal}
\end{figure*}
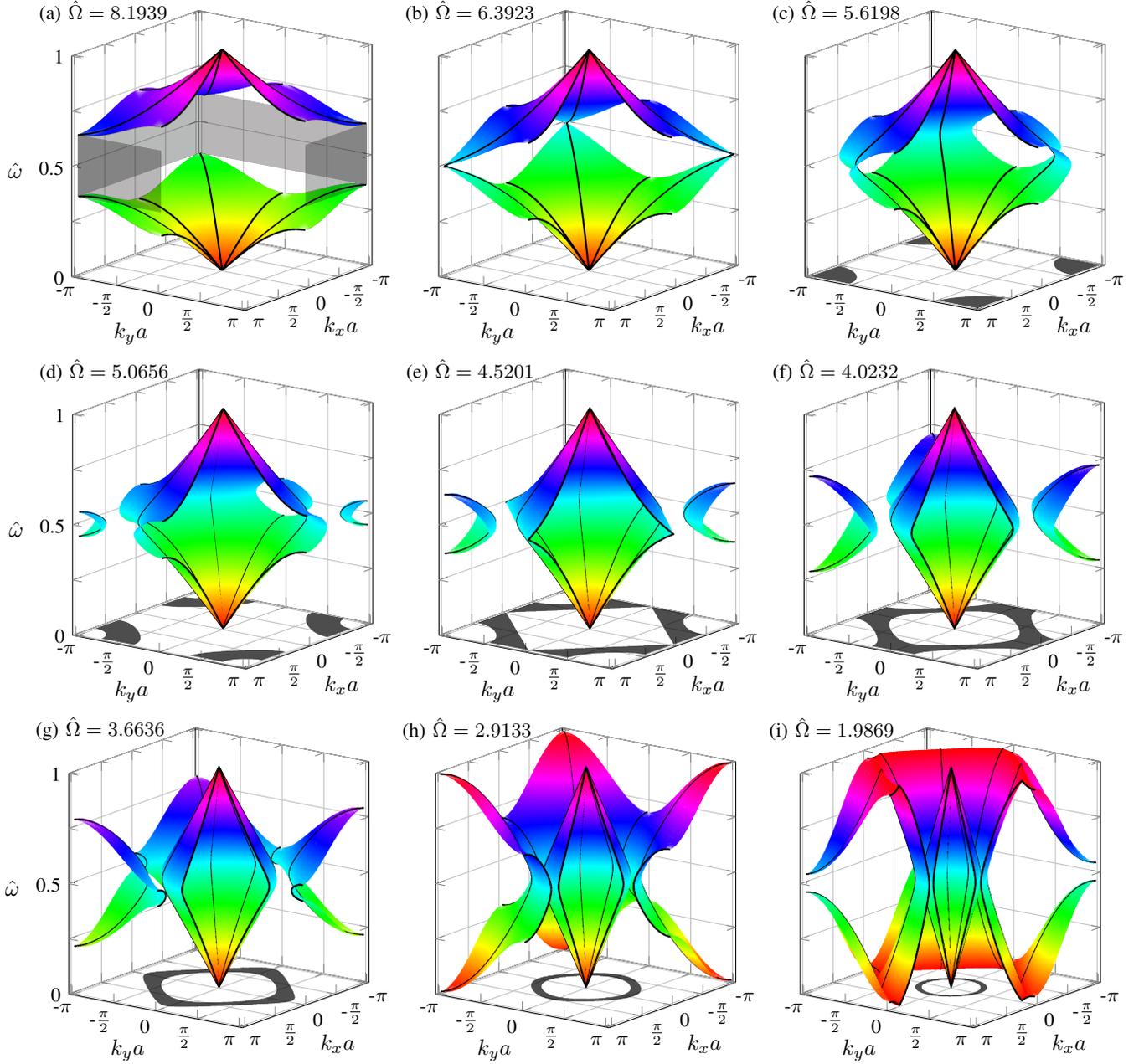

For finite modulation frequency $\Omega$, if we divide the ordinate $\omega/\omega_{0}$, in Fig.\ref{RD_Estatico2D}, by $\hat{\Omega}(=\Omega/\omega_{0})$ we get the reduced frequency $\hat{\omega}$ and this surface just represents the $\hat{\omega}_{0}^{+}$ band of the empty temporal lattice limit $m_{c}=0$. Reflecting this surface in the $\hat{\omega}=1/2$ we get the corresponding $\hat{\omega}_{1}^{-}$ band for vanishing modulation. Translating these surfaces in the vertical direction by all the integers $m$, the entire 2D+1 band structure of the empty temporal lattice can be obtained, as follows from (\ref{Eq_eigen_2D_empty_simplify}). Giving the parameter $\hat{\Omega}$ a series of values, the generalization of Fig.\ref{AtlasVacia}, for any propagation direction, is gotten.

For finite modulations $m_{c}$ we obtain the 2D+1 band structure by solving the eigenvalue equation (\ref{Eq_eigen_2D_normalizada}). For weak modulation, $m_{c}=0.36$, the ``atlas'' of Fig.\ref{Temporal} shows the qualitative changes that occur as the parameter $\hat{\Omega}$ is decreased in unequal steps from the value $8.1939$ in Fig.\ref{8temp1939} to $1.9869$ in Fig.\ref{1temp9869}. This can be considered a generalization of Fig.\ref{Atlas01} (for the red lines) for propagation in the \hkl<10> directions to propagation in an arbitrary direction in the $x-y$ plane.

\begin{figure*}[!t]
\captionsetup[subfigure]{labelformat=empty}
\centering
\subfloat[\label{dp11}]{\input{SubFigura6a}}%
\subfloat[\label{dp01}]{\input{SubFigura6b}}%
\caption{``Diábolos'' formed at the temporal BZ edge $\hat{\omega}=1/2$ and the spatial high-symmetry BZ edges $\mathbf{M}$ in (a) and $\mathbf{X}$ in (b). Note the correspondence between (a) and Fig.\ref{6temp3926} and between (b) and Fig.\ref{4temp5201} and Fig.\ref{4atls5201}. The black lines emphasize the high symmetry directions [11] and [10].}
\label{dp}
\end{figure*}
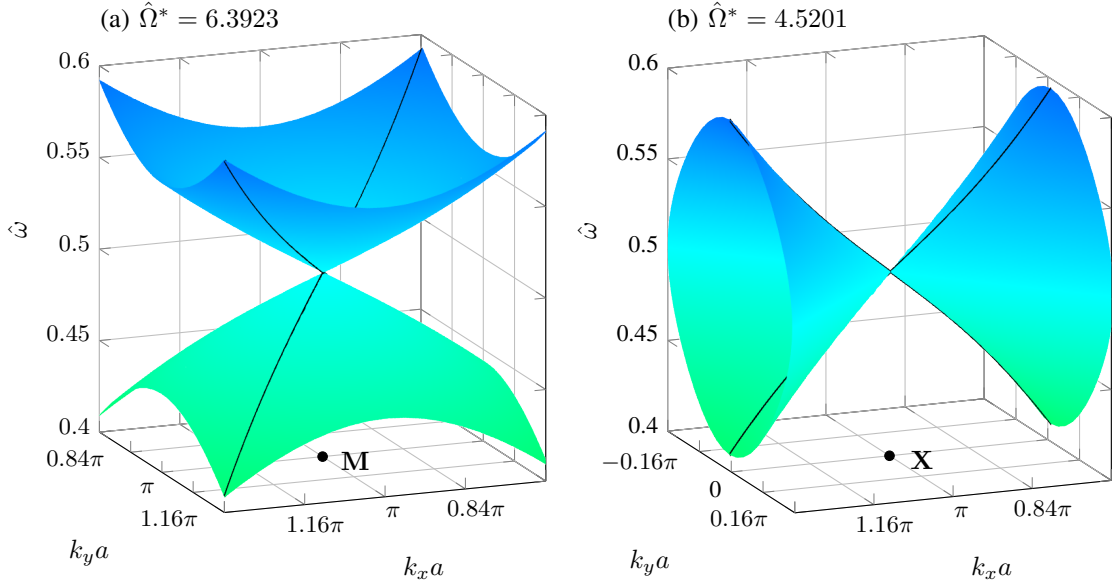

For $\hat{\Omega}>6.3923$ there is a frequency gap (for any direction $\mathbf{k}$) centered at $\hat{\omega}=1/2$ and for $\hat{\Omega}=8.1939$ the gap is of width $\sim 1/4$, as determined by the values of $\hat{\omega}$ of the + and - bands at the vertices of the BZ, see Fig.\ref{8temp1939}. For $\hat{\Omega}=6.3923$ in Fig.\ref{6temp3926}, these bands just come to touch at the vertices, thus obliterating the band gap. Between $\hat{\Omega}=6.3923$ and $\hat{\Omega}=3.6636$ (Fig.\ref{5temp6198}-\ref{3temp6636}) \emph{wave vector} forbidden bands are created; these correspond to regions of the $(k_{x}$, $k_{y})$ plane that are excluded from propagation (see shaded areas in the $\hat{\omega}=0$ plane) \emph{for any value of $\hat{\omega}$}. In Fig.\ref{5temp6198}, for $\hat{\Omega}=5.6198$ these $\mathbf{k}$-gaps have the form of small, almost circular, disks that are centered at the vertices of the spatial BZ. Lowering $\hat{\Omega}$ further, to $\hat{\Omega}=5.0656$ in Fig.\ref{5temp0656}, these disks become annular. In Fig.\ref{4temp5201}, for $\hat{\Omega}=4.5201$ these annular regions increase becoming quasi-square. At the same time, there develops a rhomb-like band centered at the $\mathbf{\Gamma }$ point and $\hat{\omega}=1/2$. This band just touches the $\mathbf{X}$ points, Fig.\ref{4temp5201}. By symmetry, in the neighboring zones there are identical ``rhombs'' touching the same zone edges from the other side. Because this structure resembles a South American toy called ``diábolo'', such shared points are sometimes refered to as ``diabolic'' \cite{berry}. The value $\hat{\Omega}=4.5201\equiv \hat{\Omega}^{*}$ marks the transition from a \emph{partial} $\hat{\omega}$-gap (for propagation in the \hkl<10> directions, see Fig.\ref{4atls5201}) for $\hat{\Omega}>\hat{\Omega}^{*}$ to complete $\mathbf{k}$-gaps for $\hat{\Omega}<\hat{\Omega}^{*}$ corresponding to a form of phase transition \cite{Alu,miller}, see Section \ref{PD}. In Fig.\ref{4temp0232}, for $\hat{\Omega}=4.0232$, the $\mathbf{k}$-gaps become distanced from the $\mathbf{M}$ points of the BZ and get nearer to the zone center. Proceeding from $\hat{\Omega}=3.6636$ in Fig.\ref{3temp6636} to $\hat{\Omega}=2.9133$ in Fig.\ref{2temp9133}, and then to $\hat{\Omega}=1.9869$ in Fig.\ref{1temp9869}, annular forbidden $\mathbf{k}$ bands form around the $\mathbf{\Gamma }$ point, gradually becoming more circular, narrower, and smaller. In Fig.\ref{2temp9133}, for $\hat{\Omega}=2.9133$, the bands at the $\mathbf{M}$ points reach the limits of the temporal BZ at $\hat{\omega}=0$ and $\hat{\omega}=1$, resembling petals of a flower. Finally in Fig.\ref{1temp9869}, with $\hat{\Omega}=1.9869$, other parts of the band also reach the same limits.

As we saw in Section \ref{10} the qualitative changes that occur for weak modulation, also occur for others modulations.

%% file: SubFigura4a.tex
\begin{tikzpicture}
\begin{axis}[width=230.0pt ,height=230.0pt,
enlargelimits=false,
axis on top,
xlabel={$k_{x}a$},
ylabel={$k_{y}a$},
ytick={-3.1416,-1.5708,0,1.5708,3.1416},
xtick={-3.1416,-1.5708,0,1.5708,3.1416},
xticklabels={-$\pi$,-$\frac 1 2 \pi$,0,$\frac 1 2 \pi$,$\pi$},
yticklabels={-$\pi$,-$\frac 1 2 \pi$,0,$\frac 1 2 \pi$,$\pi$},
]
\addplot graphics [
xmin=-3.1416 ,xmax=3.1416 ,ymin=-3.1416,ymax=3.1416,
includegraphics={trim=19.8 21.5 6.5 6.75,clip}, 
] {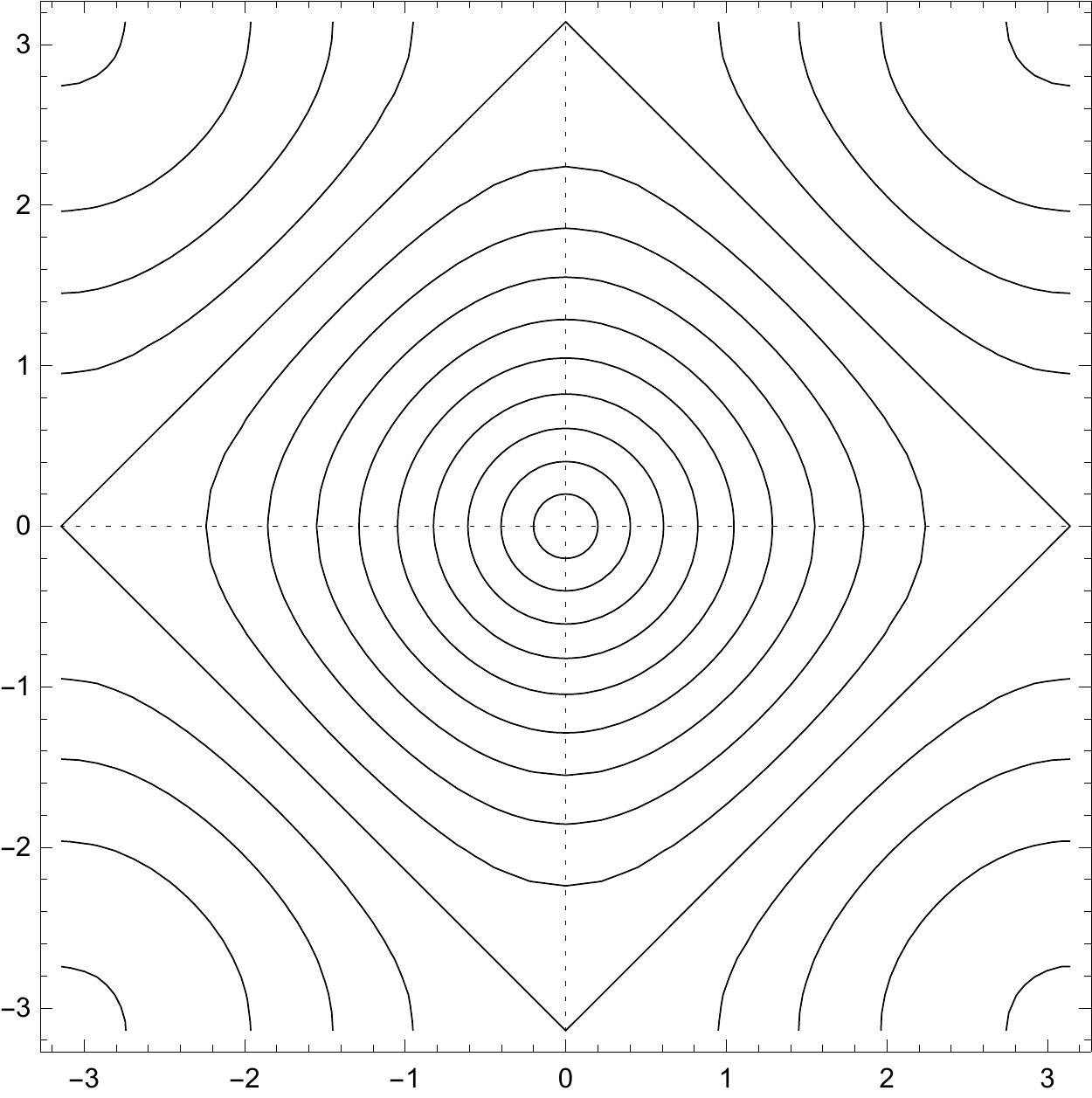};
\fill [fill=gray!50](0.9,0.1) rectangle (1.1,0.4);
\draw (1,0.25) node[black] {\small 1}; 
\fill [fill=gray!50](2.8,0.1) rectangle (3,0.4);
\draw (2.9,0.25) node[black] {\small 2};
\fill [fill=gray!50](2.6,2.45) rectangle (3.2,2.75);
\draw (2.9,2.6) node[black] {\small 2.8};
\end{axis}
\draw (0.2,6.7) node {(a)};
\end{tikzpicture} 

%% file: SubFigura4b.tex
\begin{tikzpicture}
\begin{axis}[width=230.0pt ,height=230.0pt,
zmin=0,
zmax=2.828,
grid=both,
xtick={-3.1416,-1.5708,0,1.5708,3.1416},
ytick={-3.1416,-1.5708,0,1.5708,3.1416},
xticklabels={-$\pi$ ,-$\frac 1 2 \pi$,0,$\frac 1 2 \pi$, $\pi$},
yticklabels={-$\pi$ ,-$\frac 1 2 \pi$,0,$\frac 1 2 \pi$, $\pi$},
xlabel={$k_{x}a$},
ylabel={$k_{y}a$},
zlabel={$\omega/\omega_{0}$},
]
\addplot3 graphics [
points={
(3.1415,-3.1415,2.828) => (0,492-190)
(-3.1415,-3.1415,2.828) => (272,492-0)
(-3.1416,3.1416,2.828) => (642,492-140)
(0,0,0) => (320,492-492)
},
] {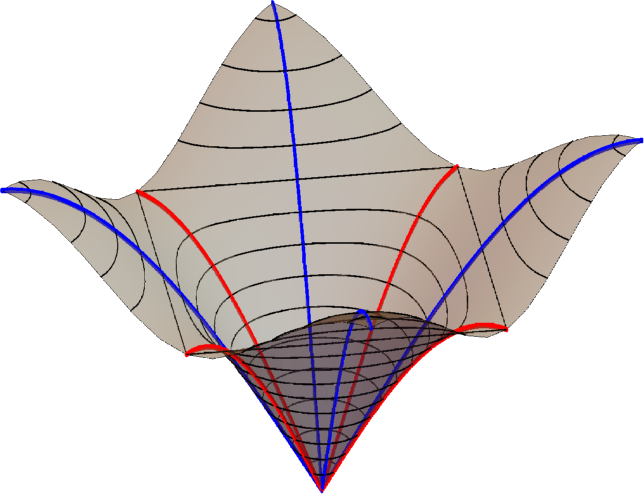};
\filldraw [black] (axis cs:0,0) circle [radius=1.8pt];
\filldraw [black] (axis cs:0,3.1416) circle [radius=1.8pt];
\filldraw [black] (axis cs:-3.1416,3.1416) circle [radius=1.8pt];
\draw [black,very thick,dashed]
(0,0) -- 
++ 
(axis direction cs:0,3.1416);
\draw [black,very thick,dashed]
(0,0) -- 
++ 
(axis direction cs:-3.1416,3.1416);
\draw [black,very thick,dashed]
(axis cs:-3.1416,3.1416) -- 
(axis cs:0,3.1416);
\node [] at (axis cs:0.5,0.5) {$\Gamma $};
\node [] at (axis cs:0.7,2.8) {X};
\node [] at (axis cs:-3.18,2.5) {M};
\end{axis}
\draw (0.2,6.7) node {(b)};
\end{tikzpicture}

%% file: Figura5.tex
\begin{tikzpicture}
\begin{axis}[width=0.35\linewidth ,height=0.35\linewidth ,
zmin=0,
zmax=1,
grid=both,
ztick={0,0.25,0.5,0.75,1},
ytick={-3.1416,-1.5708,0,1.5708,3.1416},
xtick={-3.1416,-1.5708,0,1.5708,3.1416},
xticklabels={-$\pi$,-$\frac \pi 2 $,0,$\frac \pi 2$,$\pi$},
yticklabels={-$\pi$,-$\frac \pi 2$,0,$\frac \pi 2$,$\pi$},
zticklabels={0,,0.5,,1},
xlabel={$k_{x}a$},
ylabel={$k_{y}a$},
zlabel={$\hat{\omega}$},
every axis z label/.style={
at={(ticklabel cs:0.5)},
anchor=near ticklabel,
rotate=false,
},
every axis y label/.style={
at={(ticklabel cs:0.5)},
},
every axis x label/.style={
at={(ticklabel cs:0.5)},
},
]
\addplot3 graphics [
points={
(0,0,1) => (347,617.9)
(3.1415,-3.1415,0.5) => (1,617.9-278)
(3.1416,3.1416,0) => (400,615.9-618)
(0,0,0) => (346,617.9-528.9)
(-3.1416,3.1416,1)
},
] {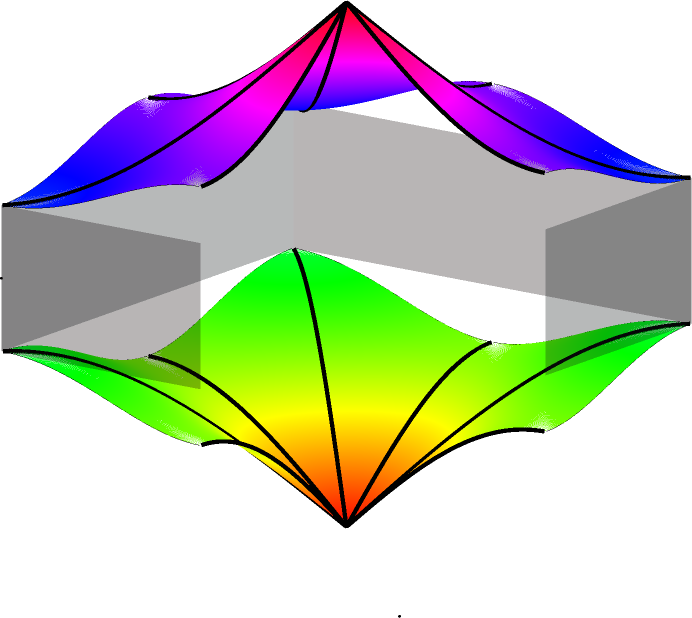};
\end{axis}
\draw (0.5,4.75) node {\small (a) $\hat{\Omega} =8.1939$}; 
\end{tikzpicture}%
\begin{tikzpicture}
\begin{axis}[width=0.35\linewidth ,height=0.35\linewidth ,
zmin=0,
zmax=1,
grid=both,
ztick={0,0.25,0.5,0.75,1},
ytick={-3.1416,-1.5708,0,1.5708,3.1416},
xtick={-3.1416,-1.5708,0,1.5708,3.1416},
xticklabels={-$\pi$,-$\frac \pi 2 $,0,$\frac \pi 2$,$\pi$},
yticklabels={-$\pi$,-$\frac \pi 2$,0,$\frac \pi 2$,$\pi$},
xlabel={$k_{x}a$},ylabel={$k_{y}a$},
zticklabels=\empty,
every axis y label/.style={
at={(ticklabel cs:0.5)},
},
every axis x label/.style={
at={(ticklabel cs:0.5)},
},
]
\addplot3 graphics [
points={
(0,0,1) => (347,617.9)
(3.1415,-3.1415,0.5) => (1,617.9-278)
(3.1416,3.1416,0) => (400,615.9-618)
(0,0,0) => (346,617.9-528.9)
(-3.1416,3.1416,1)
},
] {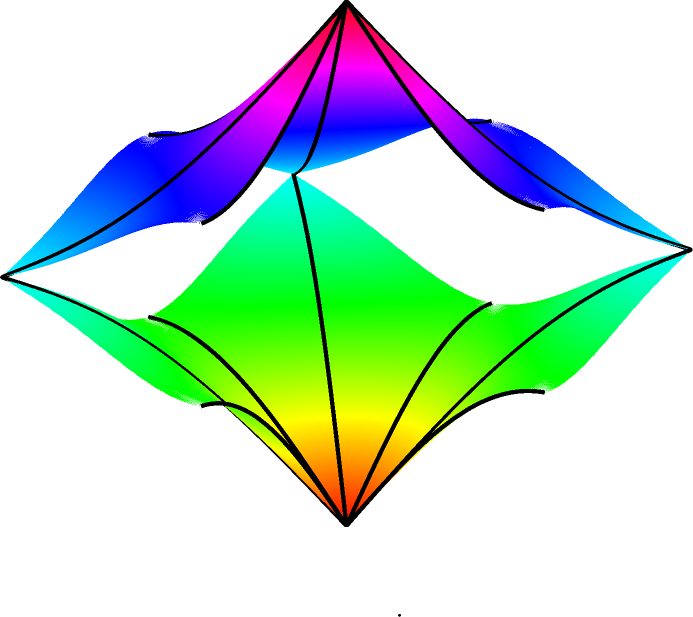};
\end{axis}
\draw (0.5,4.75) node {\small (b) $\hat{\Omega} =6.3923$}; 
\end{tikzpicture}%
\begin{tikzpicture}
\begin{axis}[width=0.35\linewidth ,height=0.35\linewidth ,
zmin=0,
zmax=1,
grid=both,
ztick={0,0.25,0.5,0.75,1},
ytick={-3.1416,-1.5708,0,1.5708,3.1416},
xtick={-3.1416,-1.5708,0,1.5708,3.1416},
xticklabels={-$\pi$,-$\frac \pi 2 $,0,$\frac \pi 2$,$\pi$},
yticklabels={-$\pi$,-$\frac \pi 2$,0,$\frac \pi 2$,$\pi$},
xlabel={$k_{x}a$},ylabel={$k_{y}a$},
zticklabels=\empty,
every axis y label/.style={
at={(ticklabel cs:0.5)},
},
every axis x label/.style={
at={(ticklabel cs:0.5)},
},
]
\addplot3 graphics [
points={
(0,0,1) => (347,617.9)
(3.1415,-3.1415,0.5) => (1,617.9-278)
(3.1416,3.1416,0) => (400,615.9-618)
(0,0,0) => (346,617.9-528.9)
(-3.1416,3.1416,1)
},
] {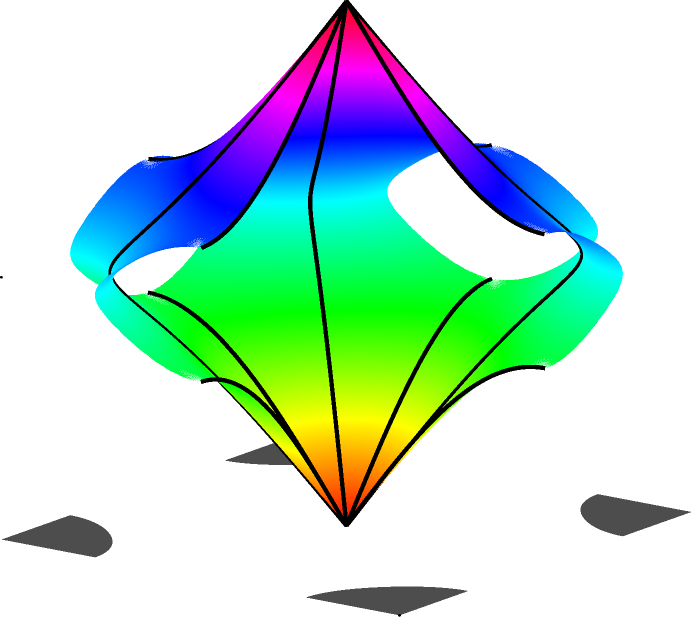};
\end{axis}
\draw (0.5,4.75) node {\small (c) $\hat{\Omega} =5.6198$}; 
\end{tikzpicture}

\begin{tikzpicture}
\begin{axis}[width=0.35\linewidth ,height=0.35\linewidth ,
zmin=0,
zmax=1,
grid=both,
ztick={0,0.25,0.5,0.75,1},
ytick={-3.1416,-1.5708,0,1.5708,3.1416},
xtick={-3.1416,-1.5708,0,1.5708,3.1416},
xticklabels={-$\pi$,-$\frac \pi 2 $,0,$\frac \pi 2$,$\pi$},
yticklabels={-$\pi$,-$\frac \pi 2$,0,$\frac \pi 2$,$\pi$},
zticklabels={0,,0.5,,1},
xlabel={$k_{x}a$},ylabel={$k_{y}a$},
zlabel={$\hat{\omega}$},
every axis z label/.style={
at={(ticklabel cs:0.5)},
anchor=near ticklabel,
rotate=false,
},
every axis y label/.style={
at={(ticklabel cs:0.5)},
},
every axis x label/.style={
at={(ticklabel cs:0.5)},
},
]
\addplot3 graphics [
points={
(0,0,1) => (347,617.9)
(3.1415,-3.1415,0.5) => (1,617.9-278)
(3.1416,3.1416,0) => (400,615.9-618)
(0,0,0) => (346,617.9-528.9)
(-3.1416,3.1416,1)
},
] {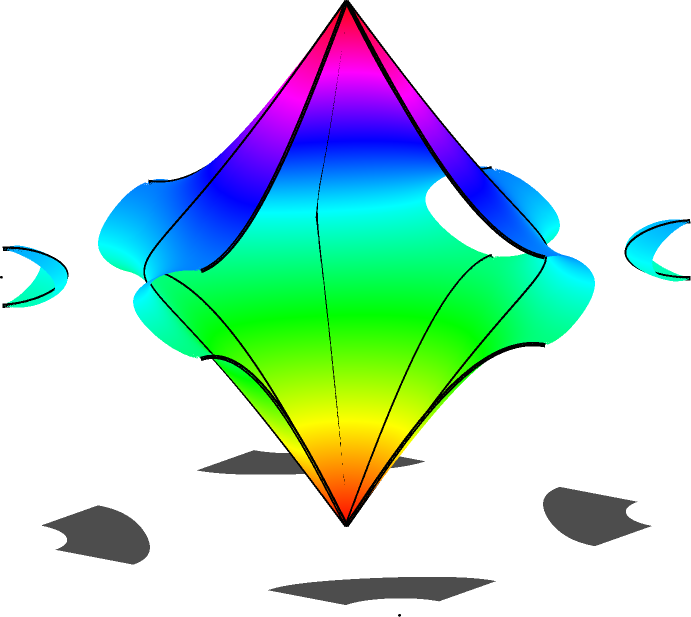};
\end{axis}
\draw (0.5,4.75) node {\small (d) $\hat{\Omega} =5.0656$}; 
\end{tikzpicture}%
\begin{tikzpicture}
\begin{axis}[width=0.35\linewidth ,height=0.35\linewidth ,
zmin=0,
zmax=1,
grid=both,
ztick={0,0.25,0.5,0.75,1},
ytick={-3.1416,-1.5708,0,1.5708,3.1416},
xtick={-3.1416,-1.5708,0,1.5708,3.1416},
xticklabels={-$\pi$,-$\frac \pi 2 $,0,$\frac \pi 2$,$\pi$},
yticklabels={-$\pi$,-$\frac \pi 2$,0,$\frac \pi 2$,$\pi$},
xlabel={$k_{x}a$},ylabel={$k_{y}a$},
zticklabels=\empty,
every axis y label/.style={
at={(ticklabel cs:0.5)},
},
every axis x label/.style={
at={(ticklabel cs:0.5)},
},
]
\addplot3 graphics [
points={
(0,0,1) => (347,617.9)
(3.1415,-3.1415,0.5) => (1,617.9-278)
(3.1416,3.1416,0) => (400,615.9-618)
(0,0,0) => (346,617.9-528.9)
(-3.1416,3.1416,1)
},
] {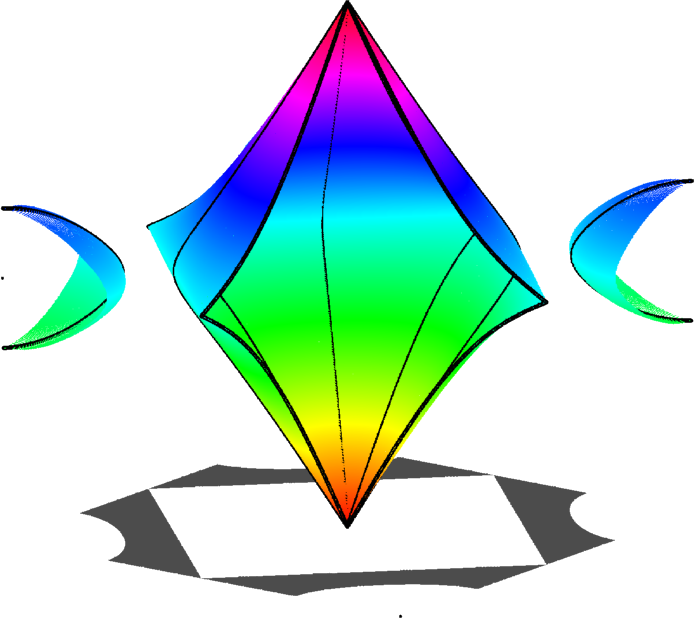};
\end{axis}
\draw (0.5,4.75) node {\small (e) $\hat{\Omega} =4.5201$}; 
\end{tikzpicture}%
\begin{tikzpicture}
\begin{axis}[width=0.35\linewidth ,height=0.35\linewidth ,
zmin=0,
zmax=1,
grid=both,
ztick={0,0.25,0.5,0.75,1},
ytick={-3.1416,-1.5708,0,1.5708,3.1416},
xtick={-3.1416,-1.5708,0,1.5708,3.1416},
xticklabels={-$\pi$,-$\frac \pi 2 $,0,$\frac \pi 2$,$\pi$},
yticklabels={-$\pi$,-$\frac \pi 2$,0,$\frac \pi 2$,$\pi$},
xlabel={$k_{x}a$},ylabel={$k_{y}a$},
zticklabels=\empty,
every axis y label/.style={
at={(ticklabel cs:0.5)},
},
every axis x label/.style={
at={(ticklabel cs:0.5)},
},
]
\addplot3 graphics [
points={
(0,0,1) => (347,617.9)
(3.1415,-3.1415,0.5) => (1,617.9-278)
(3.1416,3.1416,0) => (400,615.9-618)
(0,0,0) => (346,617.9-528.9)
(-3.1416,3.1416,1)
},
] {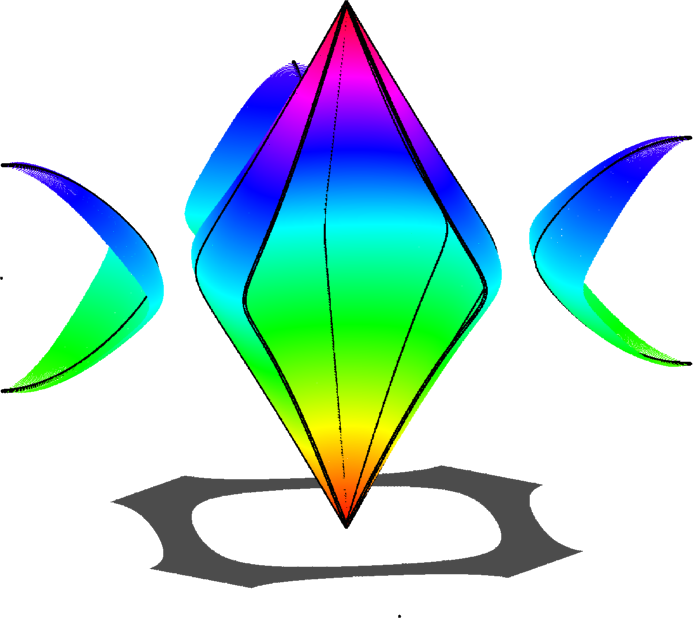};
\end{axis}
\draw (0.5,4.75) node {\small (f) $\hat{\Omega} =4.0232$}; 
\end{tikzpicture}

\begin{tikzpicture}
\begin{axis}[width=0.35\linewidth ,height=0.35\linewidth ,
zmin=0,
zmax=1,
grid=both,
ztick={0,0.25,0.5,0.75,1},
ytick={-3.1416,-1.5708,0,1.5708,3.1416},
xtick={-3.1416,-1.5708,0,1.5708,3.1416},
xticklabels={-$\pi$,-$\frac \pi 2 $,0,$\frac \pi 2$,$\pi$},
yticklabels={-$\pi$,-$\frac \pi 2$,0,$\frac \pi 2$,$\pi$},
zticklabels={0,,0.5,,1},
xlabel={$k_{x}a$},ylabel={$k_{y}a$},
zlabel={$\hat{\omega}$},
every axis z label/.style={
at={(ticklabel cs:0.5)},
anchor=near ticklabel,
rotate=false,
},
every axis y label/.style={
at={(ticklabel cs:0.5)},
},
every axis x label/.style={
at={(ticklabel cs:0.5)},
},
]
\addplot3 graphics [
points={
(0,0,1) => (347,617.9)
(3.1415,-3.1415,0.5) => (1,617.9-278)
(3.1416,3.1416,0) => (400,615.9-618)
(0,0,0) => (346,617.9-528.9)
(-3.1416,3.1416,1)
},
] {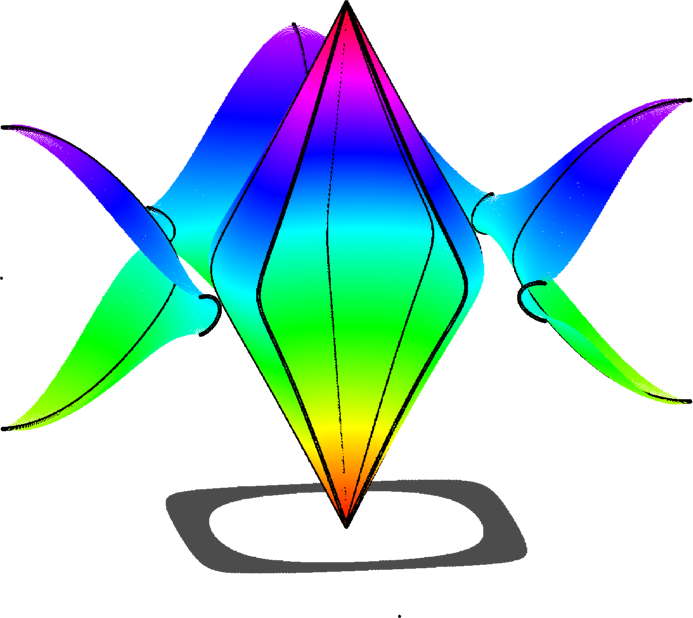};
\end{axis}
\draw (0.5,4.75) node {\small (g) $\hat{\Omega} =3.6636$}; 
\end{tikzpicture}%
\begin{tikzpicture}
\begin{axis}[width=0.35\linewidth ,height=0.35\linewidth ,
zmin=0,
zmax=1,
grid=both,
ztick={0,0.25,0.5,0.75,1},
ytick={-3.1416,-1.5708,0,1.5708,3.1416},
xtick={-3.1416,-1.5708,0,1.5708,3.1416},
xticklabels={-$\pi$,-$\frac \pi 2 $,0,$\frac \pi 2$,$\pi$},
yticklabels={-$\pi$,-$\frac \pi 2$,0,$\frac \pi 2$,$\pi$},
xlabel={$k_{x}a$},ylabel={$k_{y}a$},
zticklabels=\empty,
every axis y label/.style={
at={(ticklabel cs:0.5)},
},
every axis x label/.style={
at={(ticklabel cs:0.5)},
},
]
\addplot3 graphics [
points={
(0,0,1) => (347,617.9)
(3.1415,-3.1415,0.5) => (1,617.9-278)
(3.1416,3.1416,0) => (400,615.9-618)
(0,0,0) => (346,617.9-528.9)
(-3.1416,3.1416,1)
},
] {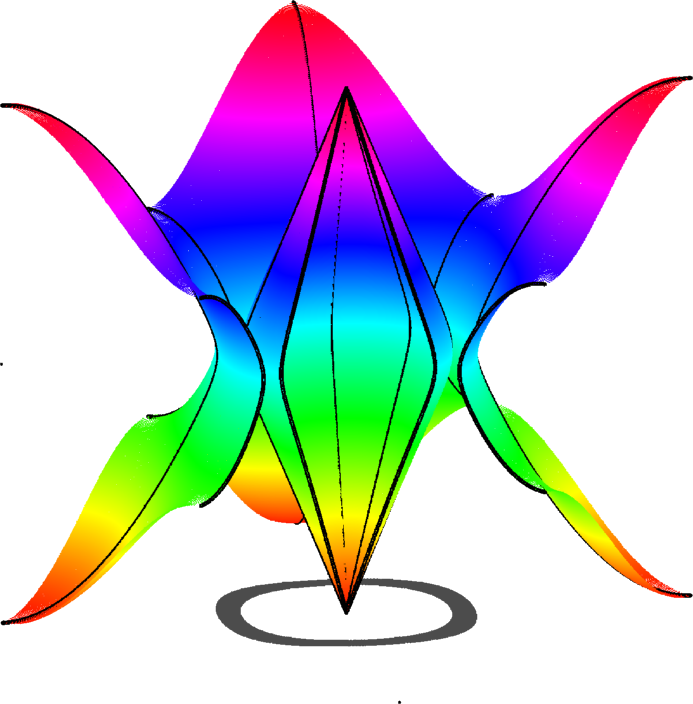};
\end{axis}
\draw (0.5,4.75) node {\small (h) $\hat{\Omega} =2.9133$}; 
\end{tikzpicture}%
\begin{tikzpicture}
\begin{axis}[width=0.35\linewidth ,height=0.35\linewidth ,
zmin=0,
zmax=1,
grid=both,
ztick={0,0.25,0.5,0.75,1},
ytick={-3.1416,-1.5708,0,1.5708,3.1416},
xtick={-3.1416,-1.5708,0,1.5708,3.1416},
xticklabels={-$\pi$,-$\frac \pi 2 $,0,$\frac \pi 2$,$\pi$},
yticklabels={-$\pi$,-$\frac \pi 2$,0,$\frac \pi 2$,$\pi$},
xlabel={$k_{x}a$},ylabel={$k_{y}a$},
zticklabels=\empty,
every axis y label/.style={
at={(ticklabel cs:0.5)},
},
every axis x label/.style={
at={(ticklabel cs:0.5)},
},
]
\addplot3 graphics [
points={
(0,0,1) => (347,617.9)
(3.1415,-3.1415,0.5) => (1,617.9-278)
(3.1416,3.1416,0) => (400,615.9-618)
(0,0,0) => (346,617.9-528.9)
(-3.1416,3.1416,1)
},
] {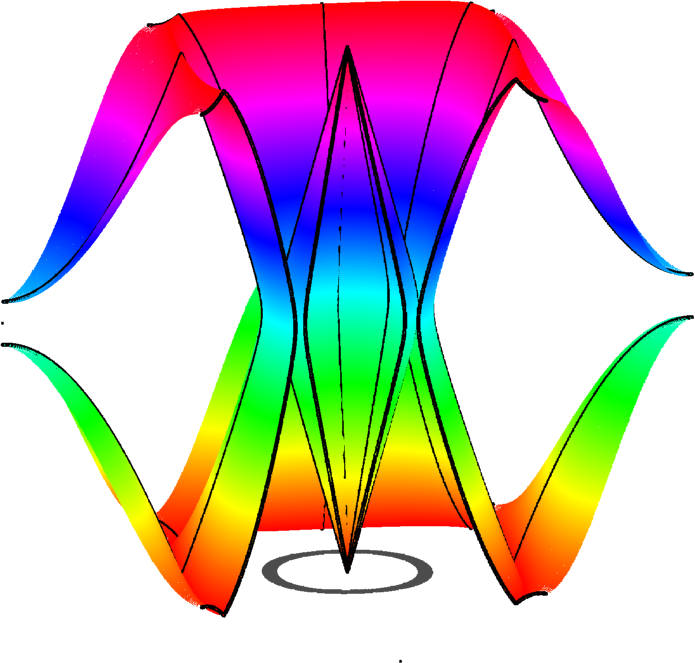};
\end{axis}
\draw (0.5,4.75) node {\small (i) $\hat{\Omega} =1.9869$}; 
\end{tikzpicture}

%% file: SubFigura6a.tex
\begin{tikzpicture}
\begin{axis}[width=230.0pt ,height=230.0pt,
enlargelimits=false,
zmin=0.4,
zmax=0.6,
xmin=2.1416,
xmax=4.1416,
ymin=2.1416,
ymax=4.1416,
grid=both,
ytick={2.1416,2.6416,3.1416,3.6416,4.1416},
xtick={2.1416,2.6416,3.1416,3.6416,4.1416},
xticklabels={,$0.84\pi$,$\pi$,$1.16\pi$,},
yticklabels={$0.84\pi$,,$\pi$,$1.16\pi$,}, 
xlabel={$k_{x}a$},
ylabel={$k_{y}a$},
zlabel={$\hat{\omega}$},
]
\addplot3 graphics [
points={
(4.1416,2.1416,0.4) => (1,708.9-608.9)
(2.1416,4.1416,0.4) => (683.9,708.9-683.9)
(2.1416,2.1416,0.6) => (491.9,708.9-0)
(3.1416,3.1416,0.5) => (342,708.9-366)
},
]{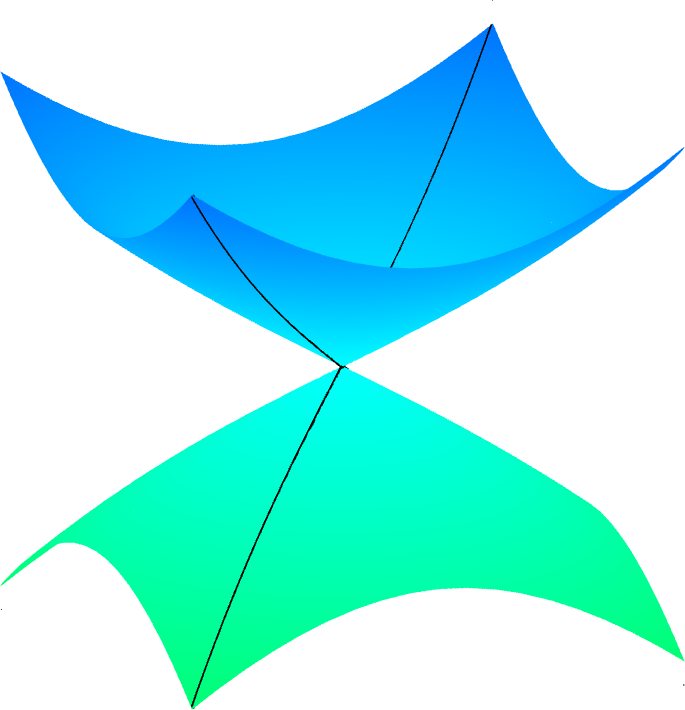};
\filldraw [black] (axis cs:3.1416,3.1416) circle [radius=1.8pt];
\node [] at (axis cs:3,3.3) {$\mathbf{M}$};
\end{axis}
\draw (1.2,6.6) node {(a) $\hat{\Omega}^{*}=6.3923$};
\end{tikzpicture}

%% file: SubFigura6b.tex
\begin{tikzpicture}
\begin{axis}[width=230.0pt ,height=230.0pt,
enlargelimits=false,
zmin=0.4,
zmax=0.6,
xmin=2.1416,
xmax=4.1416,
ymin=-1,
ymax=1,
grid=both,
ytick={-1,-0.5,0,0.5,1},
xtick={2.1416,2.6416,3.1416,3.6416,4.1416},
ztick={0.4,0.45,0.5,0.55,0.6},
xticklabels={,$0.84\pi$,$\pi$,$1.16\pi$,},
yticklabels={$-0.16\pi$,,0,$0.16\pi$,},
zticklabels={0.4,0.45,0.5,0.55,0.6},
xlabel={$k_{x}a$},
ylabel={$k_{y}a$},
zlabel={$\hat{\omega}$},
]
\addplot3 graphics [
points={
(4.1416,-1,0.4) => (0,685-610)
(2.1416,1,0.4) => (684,685-686)
(2.1416,-1,0.6) => (492,685-0)
(3.1416,0,0.5) => (344,685-368)
},
]{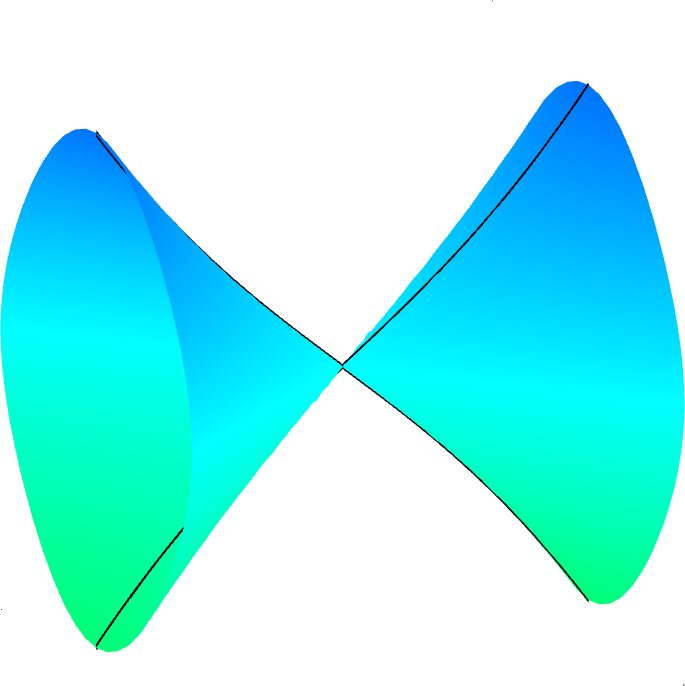};
\filldraw [black] (axis cs:3.1416,0) circle [radius=1.8pt];
\node [] at (axis cs:3,0.16) {$\mathbf{X}$};
\end{axis}
\draw (1.2,6.6) node {(b) $\hat{\Omega}^{*}=4.5201$};
\end{tikzpicture}

%% file: SecDiabolicPoints.tex
\section{\label{PD}Diabolic points.}

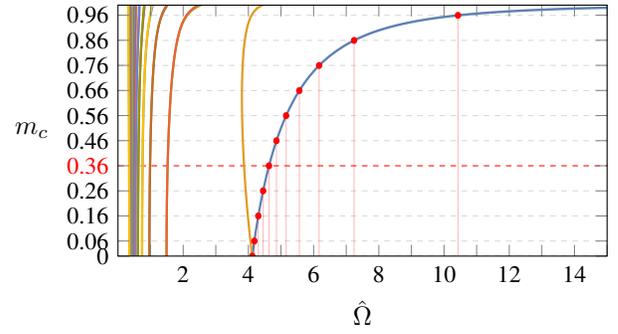
\begin{figure}[!t] 
\centering
\input{Figura7}%
\caption{Eigenvalues of (\ref{Eq_eigen_2D_normalizada}) for $\hat{\omega}=1/2$ and $(k_{x}a, k_{y}a) = (\pi, 0)$, namely the $\mathbf{X}$ point. For every value of the modulation $m_{c}$ there is an infinite number of solutions for $\hat{\Omega}$, however only the greatest values (red dots on the rightmost curve) are critical values $\hat{\Omega}^{*}$.}
\label{Omegamc}
\end{figure}

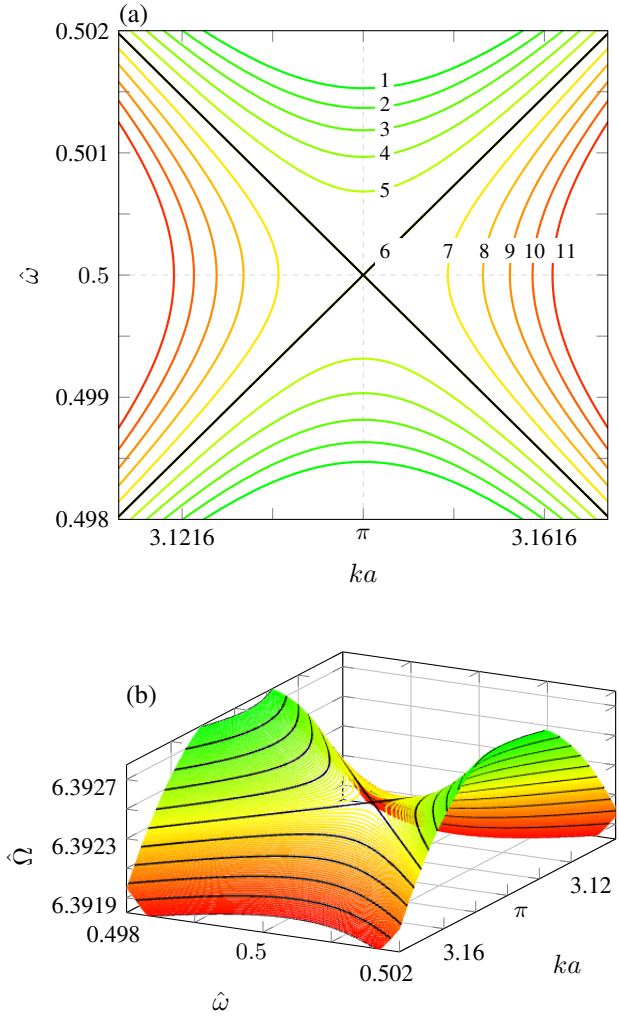
\begin{figure}[!t] 
\captionsetup[subfigure]{labelformat=empty}
\centering
\subfloat[\label{pe}]{\input{SubFigura8a}}

\subfloat[\label{silla}]{\input{SubFigura8b}}%
\caption{Transversal cut of Fig.(\ref{dp11}) in the [11] direction with zoom in the neighborhood of the diabolic point with small increments $\Delta \hat{\Omega}=7\times 10^{-5}$. The $\hat{\Omega}$ values corresponding to the 11 lines are (1)$\hat{\Omega}=6.392739$, (2)$\hat{\Omega}=6.392669$, (3)$\hat{\Omega}=6.392599$, (4)$\hat{\Omega}=6.392529$, (5)$\hat{\Omega}=6.392459$, (6)$\hat{\Omega}^{*}=6.392389$, (7)$\hat{\Omega}=6.392319$, (8)$\hat{\Omega}=6.392249$, (9)$\hat{\Omega}=6.392179$, (10)$\hat{\Omega}=6.392109$, (11)$\hat{\Omega}=6.392039$. The diabolic point is given by the intersection of the lines 6 and is obtained for $\hat{\Omega}=\hat{\Omega}^{*}=6.3923894$. The $\hat{\omega}(ka)$ lines are rendered in red for $\hat{\Omega}<\hat{\Omega}^{*}$ and in green for $\hat{\Omega}>\hat{\Omega}^{*}$. (b) Three-dimensional surface $\hat{\Omega}(\hat{\omega},ka)$ corresponding to (a). In this representation the diabolic point of (a) becomes a saddle point of the surface.}
\label{pes}
\end{figure}

We recall Fig.\ref{6temp3926} where a particular set of parameters ($m_{c}=0.36$ and $\hat{\Omega}=6.3923\equiv \hat{\Omega}^{*}$) marks a transition from a frequency gap (as in Fig.\ref{8temp1939}) to wave vector gaps (as in Fig.\ref{5temp6198}). The same behavior is, of course, displayed in neighboring BZs; the geometry in ($\hat{\omega}$, $k_{x}a$, $k_{y}a$) space is better appreciated by centering the figure at the $\mathbf{M}$-point ($k_{x}a=k_{y}a=\pi$), see Fig.\ref{dp11}. Similarly, for another specific set of prameters ($m_{c}=0.36$ and $\hat{\Omega}=4.5201\equiv \hat{\Omega}^{*}$), a transition occurs from propagation forbidden in the $\Gamma$ $\mathbf{X}$ direction (namely, a partial $\omega$-gap, Fig.\ref{5temp0656} and Fig.\ref{5atls1430}) to propagation in this direction (namely, a $k$-gap, Fig.\ref{4temp0232} and Fig.\ref{4atls0002}). Again, it is instructive to observe this behavior by a translation to the $\mathbf{X}$-point, see Fig.\ref{dp01}. In both the Fig.\ref{dp11} and Fig.\ref{dp01}, two conical surfaces touch at a high symmetry point, respectively the $\mathbf{M}$- and $\mathbf{X}$-point and at $\hat{\omega}=1/2$. These surfaces resemble a South American toy called ``diábolo'' and are, hence, sometimes refered to as ``diabolic points'' \cite{berry}. They correspond to critical values of the parameters ($m_{c}$, $\hat{\Omega}^{*}$) of our 2D+1 system \cite{Alu}.

The Fig.\ref{Temporal} and Fig.\ref{dp} suggest that diabolic points are obtained only at BZ edges in high symmetry ($\Gamma$ $\mathbf{X}$ and $\Gamma$ $\mathbf{M}$) directions and, indeed, we have not found such points for other directions of propagation. Worth noting that $\hat{\omega}=1/2$ is also a BZ edge, corresponding to the temporal periodicity. Solving (\ref{Eq_eigen_2D_normalizada}) for $\hat{\omega}=1/2$ at the $\mathbf{X}$-point, we find an infinite number of eigenvalues $\hat{\Omega}$ for every given value of $m_{c}$, but only one of these values, namely $\hat{\Omega}^{*}$, corresponds to a critical point, as can be seen in Fig.\ref{Omegamc}. In the limit $m_{c}=0$ in (\ref{Eq_eigen_2D_normalizada}) yields a very simple result for the eigenvalues, namely $\hat{\Omega}=4/n$, with $n$ being an odd integer.

For a deeper understanding of the transition involved in the neighborhood of a diabolic point, in Fig.\ref{pe} we present a vertical cross-section of Fig.\ref{dp11} in the [11] direction. While we fix $m_{c}$ at the value 0.36, $\hat{\Omega}$ varies in tiny steps of just $\Delta \hat{\Omega}=7\times 10^{-5}$. Considering the sequence of curves 1, 2, ..., 11 (corresponding to gradually decreasing values of $\hat{\Omega}$) a qualitative change in behavior occurs as we proceed from values $\hat{\Omega}>\hat{\Omega}^{*}=6.392389$ (curves 1 to 5) to values $\hat{\Omega}<\hat{\Omega}^{*}$ (curves 7 to 11). Namely, an infinitesimal decrease of $\hat{\Omega}$ causes frequency gaps to give way to wave vector gaps. The intersection of the line 6 thus represents a critical point; for $\hat{\Omega}=\hat{\Omega}^{*}$ there is neither an $\hat{\omega}$-gap nor a $k$-gap. This can be considered as a phase transition for no propagation (for $\hat{\Omega}>\hat{\Omega}^{*}$) to propagation (for $\hat{\Omega}<\hat{\Omega}^{*}$). Worth noting is that the width of the band gaps (both $\hat{\omega}$ and $k$) increase as the critical point is approached.

Fig. \ref{silla} is a three-dimensional rendition of Fig.\ref{pe} with $\hat{\Omega}$ plotted in the vertical direction. In this representation a critical point becomes a saddle point.

%% file: Figura7.tex
\begin{tikzpicture}
\begin{axis}[width=230.0pt ,height=140.0pt, 
enlargelimits=false,
axis on top,
xlabel={$\hat{\Omega}$},
ylabel={$m_{c}$},
every axis y label/.style={
at={(ticklabel cs:0.5)},
anchor=near ticklabel,
rotate=false,
},
ytick={0,0.06,0.16,0.26,0.36,0.46,0.56,0.66,0.76,0.86,0.96,1},
xtick={1,2,3,4,5,6,7,8,9,10,11,12,13,14,15},
xticklabels={ ,2, ,4, ,6, ,8, ,10, ,12, ,14, },
yticklabels={0,0.06,0.16,0.26,\textcolor{red}{0.36},0.46,0.56,0.66,0.76,0.86,0.96,},
]
\addplot graphics [
xmin=0 ,xmax=15 ,ymin=0,ymax=1,
includegraphics={trim=0 5.53 5.60 4.90,clip}, 
] {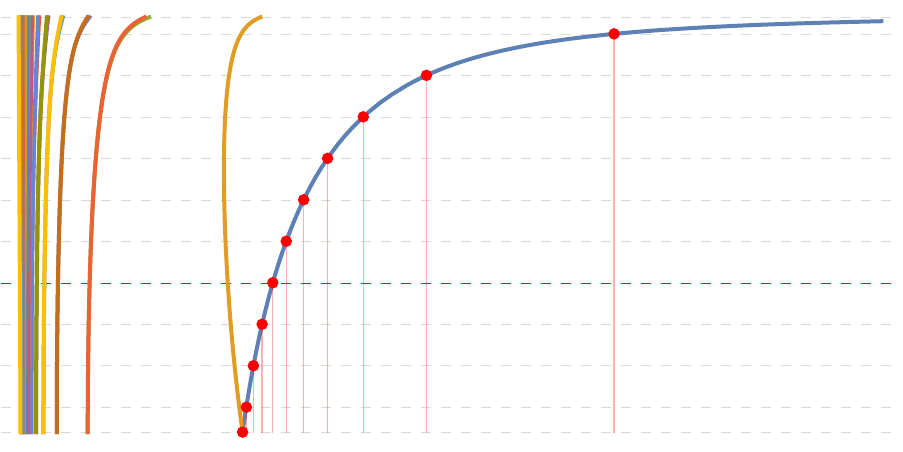};
\end{axis}
\end{tikzpicture}%

%% file: SubFigura8a.tex
\begin{tikzpicture}
\begin{axis}[width=230.0pt ,height=230.0pt,
enlargelimits=false,
axis on top,
xlabel={$ka$},
ylabel={$\hat{\omega}$},
ytick={0.498,0.4985,0.499,0.4995,0.5,0.5005,0.501,0.5015,0.502},
xtick={3.1216,3.1316,3.1416,3.1516,3.1616},
xticklabels={3.1216, ,$\pi$, ,3.1616},
yticklabels={0.498, ,0.499, ,$0.5$, ,0.501, ,0.502},
]
\addplot graphics [
xmin=3.1146 ,xmax=3.1686 ,ymin=0.498,ymax=0.502,
includegraphics={trim=0 0 0 0,clip}, 
] {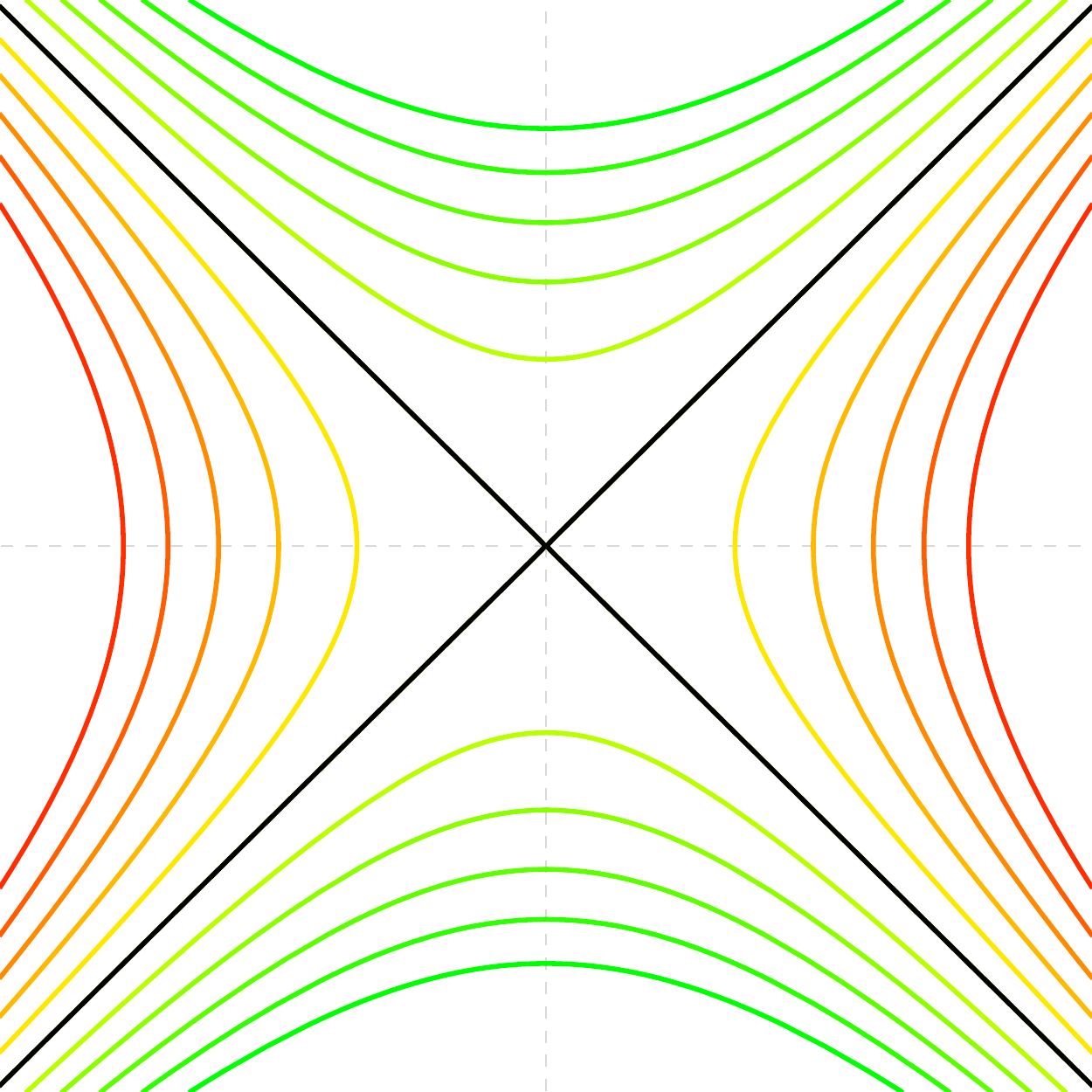};
\fill [fill=white!50](3.143,0.50005) rectangle (3.145,0.502);
\draw (3.144,0.5016) node[black] {\footnotesize 1}; 
\draw (3.144,0.5014) node[black] {\footnotesize 2};
\draw (3.144,0.5012) node[black] {\footnotesize 3};
\draw (3.144,0.5010) node[black] {\footnotesize 4};
\draw (3.144,0.5007) node[black] {\footnotesize 5};
\draw (3.144,0.5002) node[black] {\footnotesize 6};
\fill [fill=white!50](3.145,0.5001) rectangle (3.165,0.5003);
\draw (3.151,0.5002) node[black] {\footnotesize 7};
\draw (3.155,0.5002) node[black] {\footnotesize 8};
\draw (3.1578,0.5002) node[black] {\footnotesize 9};
\draw (3.1605,0.5002) node[black] {\footnotesize 10};
\draw (3.164,0.5002) node[black] {\footnotesize 11};
\end{axis}
\draw (0.2,6.7) node {(a)};
\end{tikzpicture}

%% file: SubFigura8b.tex
\begin{tikzpicture}
\begin{axis}[width=230.0pt ,height=230.0pt ,
zmin=6.3918,
zmax=6.3928,
grid=both,
ytick={0.498,0.499,0.5,0.501,0.502},
xtick={3.1216,3.1316,3.1416,3.1516,3.1616},
ztick={6.3919,6.3921,6.3923,6.3925,6.3927},
xticklabels={3.12, ,$\pi$, ,3.16},
yticklabels={0.498, ,$0.5$, ,0.502},
zticklabels={6.3919,,6.3923,,6.3927},
xlabel={$ka$},
ylabel={$\hat{\omega}$},
zlabel={$\hat{\Omega}$},
]
\addplot3 graphics [
points={
(3.1756,0.498,6.3918) => (0,260-223)
(3.1416,0.5,6.3923) => (230,260-119)
(3.1076,0.502,6.3918) => (458.9,260-154)
(3.1756,0.502,6.3918) => (256,260-260)
(3.1076,0.498,6.3928)
},
]{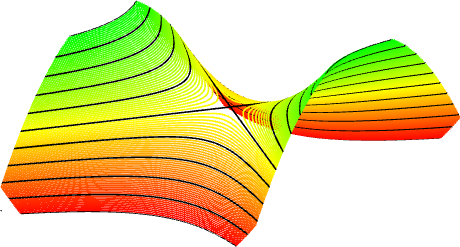};
\end{axis}
\draw (0.2,3.4) node {(b)};
\end{tikzpicture}

%% file: SecConclusion.tex
\section{\label{Conclu}Conclusions.}

In this paper we explored electromagnetic wave propagation in a system that is periodic both in space (in a plane) and in time, namely a discrete 2D low-pass TL with capacitors that are periodically modulated in tandem, see Fig.\ref{Cto_Celda_2D}. Kirchhoff's laws lead to an eigenvalue problem whose solution is an electromagnetic band structure for the frequency $\omega$ as function of the phase advances $k_{x}a$ and $k_{y}a$, see Fig.\ref{Temporal}. The surfaces $\omega(k_{x}a, k_{y}a)$ are periodic in $\omega$ with period $\Omega$ and in $k_{x}a$ and $k_{y}a$ with period $2\pi$. These exotic band structures can display forbidden $\omega$ bands, forbidden $k$ bands, both, or neither, depending on the values of two parameters: the strength of modulation $m_{c}$ of the capacitances and the reduced modulation frequency $\hat{\Omega} = \Omega/\omega_{0}$ (where $\omega_{0}$ is the LC resonance frequency of a unit cell). It turns out that, for every value of $m_{c}$ there is one critical value $\hat{\Omega}^{*}$ for the [10] direction and another critical value, namely $\sqrt{2}\hat{\Omega}^{*}$ for the [11] direction at which a kind of phase transition occurs from frequency stop bands to wave vector forbidden bands: no wave propagation for $\hat{\Omega} > \hat{\Omega}^{*}$ giving way to propagation for $\hat{\Omega} < \hat{\Omega}^{*}$, see Fig.\ref{pe}. We find such critical behavior only at the high symmetry points $\mathbf{X}$ and $\mathbf{M}$ in the 2D spatial BZ and $\omega = (1/2)\Omega$, namely a boundary of the temporal BZ. At such boundaries the $\omega(k_{x}a, k_{y}a)$ surfaces have the form of the toy ``diábolo", so that such critical points are also known as ``diabolic points". Moreover, surfaces such as $\hat{\Omega}(\omega, k_{x}a)$ exhibit a saddle point just at $\hat{\Omega} = \hat{\Omega}^{*}$, see Fig.\ref{silla}.

Our paper gives a glimpse into the rich propagation characteristics of waves in systems that are modulated periodically in time, as well as in space. It is about periodicity and symmetry in space and time and consequent critical phenomena. Besides, our system is a emerging platform that provides facility to investigate the diabolic points and their exotic phenomena. A generalization of the 2D spatial periodicity to 3D spatial periodicity is underway. Exploration of parametric resonances for finite size modulated transmission lines \cite{ZuritaResonancias}  and associated energy transfer phenomena would be of great interest. The experimental counterparts of such studies could be readily realized in the microwave and, perhaps, terahertz regime. And, we expect similar behavior to occur in modulated granular phononic crystals.